\documentclass[aps,prb,twocolumn,showpacs,showkeys,floatfix,superscriptaddress,10pt]{revtex4-1}

\usepackage{times}
\usepackage{graphicx}%
\usepackage{subfigure}
\usepackage{color}
\usepackage{bm}%
\newcommand{\orb}{f}
\newcommand{\vk} {\vec{k}}
\newcommand{\vrs} {\vec{r}^{\prime}}
\newcommand{\beq}{\begin{equation}}
\newcommand{\eeq}{\end{equation}}
\newcommand{\beqn}{\begin{eqnarray}}
\newcommand{\eeqn}{\end{eqnarray}}
\newcommand{\beqns}{\begin{eqnarray*}}
\newcommand{\eeqns}{\end{eqnarray*}}

\renewcommand{\vec}[1]{{\bf #1}}
\newcommand{\up}{\uparrow}
\newcommand{\down}{\downarrow}
\newcommand{\usepictures}[1]{#1}    %

\renewcommand{\imath}{{\rm i}}
\newcommand{\KS}{KS }
\newcommand{\HF}{HF }

\begin{document}
\title{On a Solution of the Self-Interaction Problem in Kohn-Sham Density Functional Theory}

\author{M. D\"ane}
\affiliation{Physical and Life Sciences, Lawrence Livermore National Laboratory, PO Box 808, L-372, Livermore, CA 94551}
\affiliation{Materials Science and Technology Division, Oak Ridge National Laboratory, Oak Ridge, TN 37831}
\author{A. Gonis}
\affiliation{Physical and Life Sciences, Lawrence Livermore National Laboratory, PO Box 808, L-372, Livermore, CA 94551}
\author{D. M. Nicholson}
\affiliation{Computer Science and Mathematics Division, Oak Ridge National Laboratory, Oak Ridge, TN 37831}
\author{G. M. Stocks}
\affiliation{Materials Science and Technology Division, Oak Ridge National Laboratory, Oak Ridge, TN 37831}

\begin{abstract}
We report on a methodology for the treatment of the Coulomb energy and potential in Kohn-Sham density functional theory that is free from self-interaction effects. 
Specifically,  we determine the Coulomb potential given as the functional derivative of the Coulomb energy with respect to the density,  
where the Coulomb energy is calculated explicitly in terms of the pair density of the Kohn-Sham orbitals.  This is accomplished by taking advantage
of an orthonormal and complete basis that is an explicit functional of the density that then allows for the functional differentiation of the pair density with 
respect to the density to be performed explicitly. 
This approach leads to a new formalism that provides an analytic, closed-form determination of the exchange potential.
This method is applied to one-dimensional model systems and to the atoms Helium through Krypton based on an exchange only implementation. Comparison of
our total energies (denoted SIF) to those obtained using the usual Hartree-Fock (HF) and optimized effective potential (OEP) methods reveals the hierarchy 
 $E_{\rm HF} \le E_{\rm OEP} \le E_{\rm SIF}$  that is indicative of the greater variation freedom implicit in the former two methods.

\end{abstract}



\maketitle

\section{Introduction}

Modern calculations of the electronic structure of condensed matter rest on density functional theory~\cite{DFT1,DFT3,PARR1,GROSS} (DFT), 
whose implementation is carried out within various forms based on the Kohn-Sham (KS) formulation of the theory~\cite{LDA1}. 
While there is a vast body of work that attests to the utility of this approach, it has well documented failures in predicting
electronic structure and related properties of materials where the effects of the Coulomb interaction are judged to be particularly strong. 
Such materials include, but are not limited to, 3$d$-transition metal oxides, 4$f$-electron rare earths and 5$f$-electron actinide systems.  
An important factor in this failure is the presence of the well-known unphysical
self-interaction terms in the Hartree~(classical) expression for the
Coulomb energy as well as in the exchange and correlation energy functional. 
While there have been numerous attempts~\cite{OEP0,OEP1,SIC-Perdew-Zunger,svane90,svane07,Mori-Cohen2006,PhysRevLett.89.143002,LNP2003,EXXMot,PhysRevA.14.36,KP1,KLI} to correct for
the presence of self-interaction, no fully satisfactory coherent
scheme has yet emerged\cite{mori-sanchez:201102,ruzsinszky:104102} (for a partial compendium of methods, see~\cite{List}).

Self-interaction was introduced into the theory through the original formulation of the so-called local density approximation (LDA) or local spin-density approximation (LSDA)~\cite{LDA1,PARR1,GROSS}, including their gradient corrected versions (GGAs), in which the Coulomb energy of the interacting particles is expressed within the the Hartree approximation as that of a classical charge distribution, $n({\bf r})$, with itself,
\beq
U_{\rm H}=\frac{1}{2}\int\int\frac{n({\bf r}_1)\,n({\bf r}_2)}{|{\bf r}_1-{\bf r}_2|}{\rm d}{\bf r}_1{\rm d}{\bf r}_2,
\label{UClassic.1}
\eeq
and the exchange energy is taken from the homogeneous electron gas results.
An advantage of the Hartree energy is its explicit dependence
on the density that allows the functional differentiation with respect to the density and the determination of its
contribution to the potential. 
Even though the problematic nature of the Hartree term was realized from the beginning, most of the currently used energy functionals~\cite{List} are still based on it. 
The presence of self-interaction reduces not only the reliability and convincing power of results obtained in a local approximation, but also affects the formal standing of the theory as a whole. 

Among the functional forms cited in~\cite{List}, some \cite{OEP0,OEP1,SIC-Perdew-Zunger,PhysRevLett.89.143002,LNP2003,svane90,svane07,Mori-Cohen2006,EXXMot,PhysRevA.14.36,KP1,KLI} stand out as strong candidates for the development of a general theory for a self-interaction free formulation of KS-DFT. 
Prominent among these are the self-interaction correction (SIC) method~\cite{SIC-Perdew-Zunger}, and the optimized effective potential (OEP) method~\cite{PhysRevA.14.36,OEP0,OEP1}. The former corrects for self-interaction on an orbital by orbital basis, but cannot be shown to remove the SI error completely~\cite{SIC-Perdew-Zunger}. 
The latter relies on the treatment of so-called orbital dependent functionals and attempts a functional differentiation of the Coulomb energy by means of a chain rule based on the $v$-representability of the density.   

It is well known that self-interaction does not arise when the Coulomb energy is calculated in terms of the pair density~\cite{DobsonRose1982}, a quantity that in the \KS formulation of DFT is determined through the Slater determinant of the \KS orbitals. 
However, the dependence of the pair density on the density is only implicit. 
In one implementation of the OEP method, this implicit behavior is attacked through the chain rule of functional derivatives 
which, in turn, involves the calculation of the (inverse) susceptibility of the \KS system~\cite{OEP1}. 
Alternative procedures, such as the parametrization of the potential~\cite{PhysRevLett.89.143002,wu:2498} have been formulated, where the coefficients are determined to minimize the energy. 

In this paper we provide details of a new approach that is also based the pair density, thereby 
eliminating self-interaction effects by construction, that results in particularly simple closed form expressions for the functional derivatives. 
The new formalism relies on the use of mathematical procedure, first introduced by Macke~\cite{Macke} and Harriman~\cite{Harriman} and further developed by Zumbach and Maschke~\cite{ZM}, whereby a function, e.g., the \KS orbitals, that depend $\it{implicitly}$ on the density, can be expanded in an orthonormal and complete basis of functions whose elements are  expressed  $\it{explicitly}$ in terms of the density - the so called equidensity basis. 
As will be discussed in Section~\ref{sec:Equidensity}, by making the ansatz of only considering the $\it{explicit}$ density dependence contained in the equidensity basis, 
the functional derivatives of the occupied \KS orbitals can then be obtained through term by term differentiation of the expanded forms. 
In doing this we specifically neglect any possible $\it{implicit}$  dependence of the expansion coefficients of the equidensity basis may have on the density - we will refer to this as the explicit equidensity basis (EEB) ansatz. 
The result of applying the EEB ansatz is that the final forms of the functional derivatives obtained are expressed analytically in terms of the gradients of the occupied orbitals, whose numerical evaluation is straightforward and simple to implement within existing DFT electronic structure codes. We refer to the new method as being self-interaction free (SIF) in order to emphasize the method's primary attribute. 

A short introduction of the method has been presented in previous work~\cite{Previous}. In the following pages, we provide complete details of the technical, i.e., algebraic, and computational, components of this method that show both the simplicity of the formalism as well as the power that derives from it. 
We demonstrate the characteristic features and efficacy of the new method by means of model one-dimensional calculations as well as calculations of the exchange potential and energies of the ground states of atomic systems from Helium to Krypton where it is possible to make detailed comparison with the results of 
calculations based on the well established OEP~\cite{PhysRevA.14.36,OEP0,OEP1} and Hartree-Fock (HF) methodologies. 

This is a rather long paper because of our intention of providing as complete and detailed an exposition of the method as possible. 
The paper takes the following form.

In Section~\ref{ExactKS}, we provide a derivation of the \KS equations identifying explicitly the terms corresponding to the correlation energy and potential, the exchange potential determined in a local approximation to the \KS equations, and point out the difficulties associated with self-interaction. 
Subsection~\ref{QMCorrect} states the quantum mechanically correct form of the Coulomb energy to be used in the \KS formalism. The equidensity basis, the formal and computational foundation of the methodology introduced here is set forth in Section~\ref{sec:Equidensity}.
The application of the formalism to model and realistic atomic system is presented in Section~\ref{sec:examples}. 
Differences between the method provided in this paper and others are discussed in Section~\ref{sec::oep_sif}.
Section~\ref{sec:conclusions} contains our conclusions.

In the interests of completeness, we also provide in the form of a set of extended appendices detailed derivations of key expressions resulting from our formulation that we used in calculating the results presented in the main part of the paper. 
\section{Kohn-Sham Equations}
\label{ExactKS}
In order to clarify the SIF methodology and its effectiveness in treating a self-interaction free Coulomb energy, 
we briefly review the basics of \KS density functional theory~\cite{LDA1,DFT1}. 
We consider a finite number, $N$, of electrons interacting via a Coulomb repulsion confined 
in a volume, $\Omega$, and moving under the action of an external potential, $v({\bf r})$. 
For simplicity, in the following we  consider only non-degenerate states and in much of the development we suppress the presence of spin.
Final expressions, however, are given in full spin-resolved form (see appendices). 

\subsection{Kohn-Sham Equations for Ground States}
\label{subsec::kohn-sham}
The Hamiltonian describing an interacting system of $N$ electrons in an external potential takes the usual form,
\beq
\label{Ham1}
{\hat H}^N={\hat V}+{\hat T}^N+{\hat U}^N,
\eeq
with the operators ${\hat V}$,  ${\hat T}^N$ and ${\hat U}^N$ corresponding, respectively, 
to the external field, the kinetic energy and the inter-particle
interaction (Coulomb repulsion). 
The ground-state energy of the system is given by the expectation value,
\beq
E_{\rm g}=\langle\Psi^N_{\rm g}|{\hat H}^N|\Psi^N_{\rm g}\rangle
,\quad \langle\Psi^N_{\rm g}|\Psi^N_{\rm g}\rangle=1,
\label{ExpVa}
\eeq
where $|\Psi^N_{\rm g}\rangle$ denotes the many-particle ground state of ${\hat H}^N$. 
For electrons, $|\Psi^N_{\rm g}\rangle$  leads to a wave function, $\Psi^N_{\rm g}({\bf r}_1,{\bf r}_2,\dots,{\bf r}_{N})$, that is antisymmetric with respect to interchange of the coordinates (and spins) of individual particles, 
according to the requirements of Fermi statistics. 
We use the notation, $|\Psi^N\rangle\rightarrow n({\bf r})$, and say $|\Psi^N\rangle$ leads 
to $n({\bf r})$, to denote the property, 
\beq
n({\bf r})=N\int|\Psi^N({\bf r}_1,{\bf
r}_2,\dots,{\bf r}_N)|^2{\rm d}{\bf r}_2\dots{\rm d}{\bf r}_N,
\label{Nrep.1}
\eeq
where
$n({\bf r})$ denotes the single-particle density function normalized to the total
number of particles, $N$. This property is formally equivalent to taking the expectation 
value with respect to $\Psi^N_{\rm g}({\bf r}_1,{\bf r}_2,\dots,{\bf r}_{N})$ of the single-particle number operator, 
${\hat n}({\bf r})=\psi^\dagger({\bf r})\psi({\bf r})$, where 
$\psi^\dagger$ and $\psi$  
are field creation and destruction operators for an electron at ${\bf r}$.
We now write (\ref{ExpVa}) in the form,
\beqn
E_{\rm g} & = & \int v({\bf r})n_g({\bf r}){\rm d}{\bf r}+
\langle\Psi^N_{\rm g}|\hat{T}+\hat{U}|\Psi^N_{\rm g}\rangle \nonumber \\
& = &
\underbrace{{\rm Min}}_{n(\vec{r})}  \left[
\int v({\bf r})n({\bf r}){\rm d}{\bf r}+F[n] \right]
=\underbrace{{\rm Min}}_{n(\vec{r})} 
E[n],
\label{ExpV1}
\eeqn
in terms of the constrained search functional~\cite{Levy,LiebConv},
\beqn
 F[n]&=& \underbrace{ {\rm  Min}
   }_{|\Psi\rangle\rightarrow n({\bf r})}\langle\Psi|{\hat T}^N+{\hat U}^N|\Psi\rangle. %
\label{ExpVb}
\eeqn
Given a  density, $n({\bf r})$, the constrained search examines all antisymmetric 
$N$-particle wave functions that lead to the density and delivers the state (in the absence of degeneracy) 
that produces the minimum value of  
$\langle\Psi^N|{\hat T}^N+{\hat U}^N|\Psi^N\rangle$. 
Cioslowski~\cite{Ciosl} has provided a formal procedure for generating all antisymmetric wave functions leading to $n(\vec{r})$ and  identifying that $\Psi_0^N(\vec{r}_1,\dots,\vec{r}_N)$ (in the absence of degeneracy) that determines $F[n]$. 
For $v$-representable densities, $F[n]$ gives the Hohenberg and Kohn functional~\cite{DFT1}, 
$F_{\rm HK}[n]$, and the minimizing state $\left| \Psi_0^N \right>$ coincides with $\left| \Psi_g^N \right>$.

For any other anti-symmetric state (wave function) 
$|\Phi^N\rangle\ne|\Psi^N_{\rm g}\rangle$ such that $|\Phi^N\rangle\rightarrow n({\bf r})$, we have,
\beq
F[n]\;\le\; \langle\Phi^N|{\hat T}+{\hat U}|\Phi^N\rangle,
\label{ExpVal}
\eeq
so that the exact ground-state energy, $E_{\rm g}$, forms a lower bound
of the expectation values of the Hamiltonian with respect to antisymmetric $N$-particle states, 
$|\Phi^N\rangle\rightarrow n({\bf r})$.

As in the initial formulation of DFT by Kohn and Sham~\cite{LDA1}, we postulate the existence of 
a fictitious non-interacting $N$-particle system described by the Hamiltonian,
\beq
{\hat H}^N_s={\hat V}_s+{\hat T}^N,
\label{NonH.1}
\eeq
under the action of an external potential, ${\hat V}_s$, whose ground-state density is identical 
to the density of the interacting system described by ${\hat H}^N$. 
In analogy with (\ref{ExpVb}), we define the constrained search functional,
\beqn
T_s[n]&=& \underbrace{ {\rm Min}
   }_{|\Phi^N\rangle\rightarrow n({\bf r})}\langle\Phi^N|{\hat T}^N|\Phi^N\rangle. %
\label{Tn1}
\eeqn
In the absence of degeneracy, the minimizing $|\Phi_s^N\rangle$ is a single Slater determinant~\cite{LiebConv} (denoted by the subscript $s$) 
of order $N$.

Because generally, $|\Phi_s^N\rangle\ne|\Psi^N_{\rm g}\rangle$, we have,
\beqn
E_{s} %
& = & \int v({\bf r})n({\bf r}){\rm d}{\bf r}+F_s[n]\;\ge\;E_{\rm g},
\label{ExpV1p}
\eeqn
where
\beq
\label{Fs.1}
F_s[n]= \langle\Phi_s^N|{\hat T}^N+{\hat U}^N|\Phi_s^N\rangle\ge\;F[n].
\eeq
Defining
the quantity,
$E_{c}[n]=F[n]-F_s[n]$, and adding and subtracting $F_s[n]$ to $E[n]$ in 
Eq. (\ref{ExpV1}), we obtain,
\beq
E[n]= \int v({\bf r})n({\bf r}){\rm d}{\bf r}+ F_s[n]+ E_c[n].
\label{ExpV1gs}
\eeq
The quantity $ E_c[n]$ is referred to as the correlation energy. 
We can view the last expression as a means of determining
$E_{\rm g}$ through the states of the non-interacting system,
$|\Phi^N_s\rangle$, given the functional difference, $E_{c}[n]$.

The Slater determinant, $|\Phi^N_s \rangle$, is obtained from
the solutions of a single-particle Schr\"odinger equation, 
which then also defines the potential, $v_s({\bf r})$,
\beq
\left[-\frac{1}{2}\nabla^2+v_s({\bf r})\right] f_i({\bf r})=\epsilon_if_i({\bf r}).
\label{phii}
\eeq
From (\ref{Nrep.1}) it follows that,
\beq
n({\bf r})=\sum_{j=1}^N \left| f_j({\bf r})\right|^2,
\label{Slatern.1}
\eeq
where the orbitals, $f_j({\bf r})$, correspond to the $N$ eigenvalues of Eq.~(\ref{phii}) 
that lie the lowest in energy. With respect to the same states, we also define the pair density for the non-interacting system~\cite{PARR1},
\begin{eqnarray}
&& n_s({\bf r}_1,{\bf r}_2) \nonumber \\
&=& {N \choose 2} \int   \left| \Phi^N({\bf r}_1,{\bf r}_2,\dots,{\bf r}_N)\right|^2{\rm d}{\bf r}_3\dots{\rm d}{\bf r}_N \\
&=& \frac{1}{4}\sum_{i,j}^N|f_i({\bf r}_1)f_j({\bf r}_2)-f_j({\bf r}_1)f_i({\bf r}_2)|^2. \label{SlaterPair.1} 
\end{eqnarray}

Now, the functional $F_s[n]$ takes the form,
\beqn
\nonumber F_s[n]&=& \sum_i\int{\rm d}{\bf r}f_i^*({\bf
r})\left[-\frac{1}{2}\nabla^2_{\bf r}\right]f_i({\bf r}) \\ \nonumber
&+&
\int{\rm d}{\bf r}_1\int{\rm d}{\bf r}_2\frac{n_s({\bf
r}_1,{\bf r}_2)}{|{\bf r}_1-{\bf r}_2|}.\\
\label{Fs}
\eeqn
The form of $F_s[n]$ is expressed in terms of the fully exchanged
two-particle density and is by construction free of self-interaction
effects. From Eq.~(\ref{phii}) we obtain the expectation value of the kinetic
energy operator,
\beqn
T_s [n]
& = &\sum_i\int{\rm d}{\bf r} \, f_i^*({\bf r}) \left[ -\frac{1}{2}\nabla^2_{\bf r}\right] f_i({\bf r}) \nonumber \\ 
& = & \sum_i^N \epsilon_i \int{\rm d}{\bf r} \,f_i^*({\bf r})
     f_i({\bf r})
       -  \int{\rm d}\,{\bf r}\,v_s({\bf r})\,n({\bf r})  \nonumber \\ 
&=& \sum_i^N \epsilon_i -  \int{\rm d}{\bf r}\,v_s({\bf r})\,n({\bf r}),
\label{ExpDel}
\eeqn
where a star denotes the complex conjugate of a quantity. 

Using these expressions, we can write
\beqn
\label{EXCE}
&& \hspace{-1cm} E_c[n]   = F[n]-F_s[n] \nonumber \\
&=& (T[n]-T_s[n]) 
\nonumber \\ &&
 + 
\int{\rm d}{\bf r}_1\int{\rm d}{\bf r}_2\,\frac{
n({\bf r}_1,{\bf r}_2;[n])- n_s({\bf r}_1,{\bf r}_2;[n])}{|{\bf r}_1-{\bf
r}_2|},
\eeqn
where $T$ denotes the exact expectation value of the kinetic energy
of the interacting system and $n({\bf r}_1,{\bf r}_2)$ is the corresponding exact two-particle
density.

Using the stationarity property of the \KS energy for the ground state with respect to the
density\cite{DFT1,DFT3,PARR1,GROSS}, $\left. \frac{\delta E_{\rm g}[n]}{\delta n({\bf r})}\right|_{n=n_{g}}=0$ and $\frac{\delta T_s[n]}{\delta n(\bf r)}=-v_s({\bf r})+c$ \cite{PARR1,GROSS} 
(where in the following the constant $c$ is supressed), we obtain
the requirement at the ground state,
\beq
\label{Vs1}
v_s({\bf r}) = v({\bf r})+ \left[  
\int{\rm d}{\bf r}_1\int{\rm
d}{\bf r}_2\frac{\frac{\delta n_s({\bf r}_1,{\bf r}_2)}{\delta n(\bf
r)}}{|{\bf r}_1-{\bf r}_2|}+v_{c}({\bf r}) \right]_{n=n_g},
\eeq
where the terms in the square brackets are the sum of the Coulomb potential and the correlation potential, 
\beq
v_{c}({\bf r})=\frac{\delta E_c[n]}{\delta n({\bf r})},
\label{XCpot}
\eeq
completing the expression for $v_s$ to be used in Eq.~(\ref{phii}).  Equation~(\ref{Vs1}) gives a unique \KS potential, leaving no freedom in its determination: It is the unique (generally within a constant) potential that leads to the density. 

The quantity $E_c[n]$ appearing in the previous discussion
is exact, but generally unknown. For the case of systems with infinite numbers of electrons, 
a correlation energy functional that is consistent with the second Hohenberg-Kohn theorem in yielding energies that are no lower than their 
experimental counterparts can be constructed based on the properties of the homogeneous 
electron gas (jellium), as developed in the work of Ceperley and Alder~\cite{Cep}. 
For finite systems, such a consistent construction of the correlation energy is still lacking. 
In the following, we set this term equal to zero and concentrate on what is known as the 
exchange only form of the \KS functional (the nomenclature to become clear below), 
that takes the form,
\beq
{\tilde E}[n]= \int v({\bf r})n({\bf r}){\rm d}{\bf r}+
F_s[n]\ge\; E_{\rm g}.
\label{ExpV1gsa}
\eeq

Our aim is to develop a methodology that, with $v_{c}({\bf r})$ set equal to zero, determines the potential (\KS potential), 
$v_s({\bf r})$, within the exchange-only mode of implementation of the theory, through the derivative of the Coulomb energy with respect to the density including the exchange term. 

\subsection{Hartree Approximation of the Coulomb Energy}
\label{Difficulty}
In the Hartree approximation without self-interaction corrections~\cite{LDA1}, the Coulomb energy of an interacting 
$N$-particle system, e.g., an atom, is approximated by the classical expression,
\beq
U_{\rm H}=\frac{1}{2}\int\frac{n({\bf r}_1)\,n({\bf r}_2)}{|{\bf r}_1-{\bf r}_2|}\,{\rm d}{\bf r}_1{\rm d}{\bf r}_2,
\label{Ucl1}
\eeq
where $n({\bf r})$ denotes the ground-state density of the interacting system, 
a form that lends itself immediately to functional differentiation with respect to $n({\bf r})$, (the density appears explicitly inside the integral). 

Computational ease, however, comes at a high price: the product of densities allows the simultaneous 
occupation of the same orbital by a single electron, in violation of the Pauli exclusion principle. 
Alternatively, an electron at position ${\bf r}_1$ is allowed to interact with itself at ${\bf r}_2$, 
leading to a clearly unphysical self-interaction effect.
\subsection{Coulomb Exact-Exchange Kohn-Sham Functional}
\label{QMCorrect}
Within the \KS formulation, the quantum-mechanically correct expression of the Coulomb repulsion energy of an 
non-interacting $N$-particle system takes the form~\cite{PARR1,DobsonRose1982},
\beq
U_{\rm QM}=
\int\frac{ n_s({\bf r}_1,{\bf r}_2)}{|{\bf r}_1-{\bf r}_2|}{\rm d}{\bf r}_1{\rm d}{\bf r}_2,
\label{QMU1}
\eeq
where $ n_s({\bf r}_1,{\bf r}_2)$ is the pair density obtained from the \KS orbitals. 
In general, it is convenient to define the exchange term,
\beq
J_s({\bf r}_1,{\bf r}_2)=2  n_s({\bf r}_1,{\bf r}_2)-n({\bf r}_1)n({\bf r}_2),
\label{ExcK.1}
\eeq
and a corresponding exchange energy,
\beq
E_x[n]=\frac{1}{2}\int\frac{ J_s({\bf r}_1,{\bf r}_2)}{|{\bf r}_1-{\bf r}_2|}{\rm d}{\bf r}_1{\rm d}{\bf r}_2.
\label{QMUK1a}
\eeq

In this case, with $f_j(\vec{r})$ denoting an orbital in the \KS determinant,  
the pair density takes the form ($\sigma$ denotes spin), 
\begin{eqnarray}
&& n_s({\bf r}_1,{\bf r}_2) \nonumber \\
&=& \frac{1}{4}\sum_{i,j}\left| f_i({\bf r}_1) \, f_j({\bf r}_2) - f_j({\bf r}_1)\,f_i({\bf r}_2)\right|^2 \label{PirDEns1} \\
&=&\frac{1}{2}\sum_{i,j} \bigg[ 
\orb_{i}^{*}(\vec{r}_1)  \orb_{j}^{*}(\vec{r}_2) \orb_{i}(\vec{r}_1) \orb_{j}(\vec{r}_2)
\nonumber \\ &&
- \orb_{i}^{*}(\vec{r}_1)  \orb_{j}^{*}(\vec{r}_2) \orb_{j}(\vec{r}_1) \orb_{i}(\vec{r}_2)
\,\delta_{\sigma_i,\sigma_j}
    \bigg],
\end{eqnarray}
and the exchange term has the form,
\begin{eqnarray}
&&J_s({\bf r}_1,{\bf r}_2)= \nonumber \\
&&-\sum_{i,j} 
\bigg[
 \orb_{i}^{*}(\vec{r}_1)  \orb_{j}^{*}(\vec{r}_2) \orb_{j}(\vec{r}_1) \orb_{i}(\vec{r}_2)
\,\delta_{\sigma_i,\sigma_j}
    \bigg].
\label{SlaterK.1}
\end{eqnarray}
Finally, the exchange contribution to the single-particle potential arising from the 
Coulomb energy, called the exchange potential $v_{\rm x}$, can be written as,
\begin{eqnarray}
v_{\rm x}({\bf r})
=\frac{\delta E_x[n]}{\delta n({\bf r})},
\label{Vx.1}
\end{eqnarray}
in which the functional dependence on the density is explicitly indicated. 

Written in terms of the expression in Eq.~(\ref{QMU1}), the energy functional to be minimized 
in the exchange-only form of the \KS formulation of ground-state DFT takes the form,
\beq
\tilde{E}[n]=\int v({\bf r})n({\bf r}){\rm d}{\bf r}+T_{\rm s}[n]+U_{\rm H}[n]+E_x[n].
\label{En.1}
\eeq

This form exhibits clearly the difficulty that has been encountered in attempts to implement 
the \KS formulation of DFT. 
While the functional differentiation of the Hartree term with respect to the density is 
straightforward (explicit dependence on density under the integral),  the orbitals that enter the definition of the exchange term are only 
implicitly dependent on the density defying immediate differentiation with respect to $n({\bf r})$ by analytic means. 
Even a brief survey of the many forms proposed for bypassing this seemingly impossible task 
would be exceedingly lengthy and take us too far afield of our intended purpose. 
At the same time, a few comments are in order.

The OEP method~\cite{OEP0,OEP1} is the best known and 
most widely used procedure for the solution of the \KS equations in the presence 
of exact exchange. 
The functional differentiation of the Coulomb energy in reconstructing the potential within the OEP method requires some computational effort, 
usually involving the calculation of the  inverse susceptibility and ultimately requiring the solution of an integral equation. 
Alternatively, the potential can be parametrized~\cite{PhysRevLett.89.143002,wu:2498} where parameters are determined in order to minimize the energy. 
The present SIF method avoids the need to solve the OEP equations, 
by using an 
analytic treatment in obtaining the functional derivative of the Coulomb energy with respect to the density. 
This is accomplished through the use of the equidensity basis formalism, introduced by Macke~\cite{Macke} and Harriman~\cite{Harriman}.  
In addition to orthonormality, Zumbach and Maschke~\cite{ZM} proved  that the basis is complete. 
The expansion of the orbitals in this basis brings out the explicit dependence on the density allowing the straightforward differentiation using the product rule.
Expressed in this way it is clear that the total variational freedom that is present in the method is that contained in the explicit density dependence in the basis. 
Henceforth we shall refer to this procedure as the explicit equidensity basis~(EEB) ansatz  in order to distinguish it from the more conventional OEP, which may well have 
additional variational freedom beyond that captured by this ansatz. Although the existence of the basis has been known for some time, as far as we are aware the use 
proposed here has not been attempted and as a consequence no results that compare the efficacy of this approach, compared say to the OEP, currently exist.

\section{Equidensity Basis}
\label{sec:Equidensity}
The calculation of the Coulomb potential in Eq.~(\ref{Vs1}) requires the functional derivative of the non-interacting pair density $n_s({\bf r}_1,{\bf r}_2)$ with respect 
to the (spin) density $n^{\sigma}({\bf r})$. 
This can be performed analytically by expressing the former in terms of an expansion that 
exhibits explicitly the single-particle density. 
Namely, for each orbital, $f_j^{\sigma}({\bf r})$, we write,
\begin{eqnarray}
f_j^{\sigma}({\bf r})&=&\sum_{\bf k}a_{\bf k}^{j,\sigma}\; \phi_{\bf k}^{\sigma}({\bf r},[n^{\sigma}]), \label{ZM:orbitalExpansion} \\
\phi_{\bf k}^{\sigma}({\bf r},[n^{\sigma}])&=&\sqrt{\frac{n^{\sigma}({\bf r})}{N^{\sigma}}}\; {\rm exp}\{ \imath \,{\bf k}\cdot{\bf R^{\sigma}}({\bf r},[n^{\sigma}]) \,\},
\label{ZM:basis}
\end{eqnarray}
where $\phi_{\bf k}^{\sigma}({\bf r},[n^{\sigma}])$ are the elements of an orthonormal and complete basis~\cite{Macke,Harriman,ZM}, {\it the equidensity basis}, 
where ${\bf k}=(k_x,k_y,k_z)$ denote a set of three signed integers, the $a^{j,\sigma}_{\vec{k}}$ are expansion coefficients, and the vector ${\bf R}^{\sigma}({\bf r},[n^{\sigma}])$ is defined by the expressions,
\begin{eqnarray}
  R_1^{\sigma}(x,y,z,[n^{\sigma}])&=& \frac{2\,\pi}{N^{\sigma}(y,z,[n^{\sigma}])} \int_{-\infty}^{x} \hspace{-.3cm} {\rm d}x^{\prime} \, n^{\sigma}(x^{\prime},y,z             ) \nonumber \\%
  R_2^{\sigma}(y,z,[n^{\sigma}])  &=& \frac{2\,\pi}{N^{\sigma}(z,[n^{\sigma}])}   \int_{-\infty}^{y} \hspace{-.3cm} {\rm d}y^{\prime} \, N^{\sigma}(y^{\prime},z,[n^{\sigma}])   \nonumber \\%
  R_3^{\sigma}(z,[n^{\sigma}])    &=& \frac{2\,\pi}{N^{\sigma}[n^{\sigma}]}       \int_{-\infty}^{z} \hspace{-.3cm} {\rm d}z^{\prime} \, N^{\sigma}(z^{\prime},[n^{\sigma}])     \label{ZM:R123}            %
\end{eqnarray}
with
\begin{eqnarray}
 N^{\sigma}(y,z,[n^{\sigma}])&=& \int_{-\infty}^{\infty} {\rm d}x^{\prime} \, n^{\sigma}(x^{\prime},y,z             ) \nonumber \\ %
 N^{\sigma}(z,[n^{\sigma}])  &=& \int_{-\infty}^{\infty} {\rm d}y^{\prime} \, N^{\sigma}(y^{\prime},z,[n^{\sigma}])   \nonumber \\ %
 N^{\sigma}[n^{\sigma}]    &=& \int_{-\infty}^{\infty} {\rm d}z^{\prime} \, N^{\sigma}(z^{\prime},[n^{\sigma}])                     \label{ZM:def:NNN}
\end{eqnarray}
where $0\le {R_1},\;{R_2},\;{R_3}\le 2\pi$. 
The transformation, ${\bf r}\rightarrow{\bf R}$, maps three-dimensional coordinate space onto 
the volume of a cube of side $2\pi$ with the points at infinity mapped onto the surface of the volume. 
Note that the $R$ are explicit functionals of the density. 
This particular choice of ${R_1,R_2,R_3}$ is not unique, with other choices, e.g. permuting coordinates or other coordinate systems, being 
possible~\cite{ZM_diff_basis} as well. 
In the following we use the definitions given in equations~(\ref{ZM:R123}) and (\ref{ZM:def:NNN}). \\
The coefficients, $ a_{\bf k}^{j,\sigma}$, are given as the overlap integrals,
\beq
a_{\bf k}^{j,\sigma}[n]=\int f_j^\sigma({\bf r,[n]})\; \phi_{\bf k}^{\sigma *}({\bf r},[n])\;\,{\rm d}{\bf r}.
\label{Coeff1}
\eeq
The functional derivative of the Coulomb energy can now be carried out through the differentiation of the orbitals under the integral sign in Eq.~(\ref{SlaterK.1}). The procedure is as follows: We replace the orbitals in the exchange term with their expansions in terms of the equidensity basis, and use the property of functional derivatives,
\beq
\frac{\delta n^{\sigma^{\prime\prime}}({\bf r}^\prime)}{\delta n^{\sigma}({\bf r})}=\delta_{\sigma,\sigma^{\prime\prime}} \,\delta ({\bf r}-{\bf r}^\prime),
\label{Fdd1}
\eeq
to perform the functional differentiation of the exchange term. 
It is clear that any density can be used to construct the equidensity basis, and the 
corresponding expansion of the \KS orbitals. 
However, only the density corresponding to the \KS orbitals is useful to construct the equidensity basis. 
This choice allows a direct functional differentiation of the basis with respect to the density at each 
iteration step.

\subsection{Coulomb Potential Calculation}

In general, the Coulomb potential is given by the functional derivative of the Coulomb energy with respect to the density,
\begin{eqnarray}
 v^{(U)}(\vrs)&=&\frac{\delta}{\delta n(\vrs)}\;\left( 
 \int\int {\rm d}\vec{r}_1 \, {\rm d} \vec{r}_2 \; U(\vec{r}_1,\vec{r}_2)\, n_s(\vec{r}_1,\vec{r}_2) \right) \nonumber \\
  &=& 
\int\int {\rm d}\vec{r}_1 \, {\rm d} \vec{r}_2 \; U(\vec{r}_1,\vec{r}_2) \, \left[ \frac{\delta \,n_s(\vec{r}_1,\vec{r}_2) }{\delta n(\vrs)}   \right].
  \label{ZM:eq:potential_u}
\end{eqnarray}
For three-dimensional systems the interaction $U$ is given by the expression,
\begin{equation}
 U(\vec{r}_1,\vec{r}_2) = \frac{1}{\left| \vec{r}_1-\vec{r}_2  \right| } \label{ZM:def3D:Ur1r2},
\end{equation}
(where clearly $U$ has no functional dependence on the density).

We split the pair density into two parts, the well known Hartree and the exchange contributions (with $i$ in $\orb_i$ a compound index that includes spin),
\begin{eqnarray}
    &&\hspace{-1.3cm} n_s(\vec{r}_1,\vec{r}_2) \nonumber \\&=& %
  \frac{1}{2}
   \sum_{ij, i\ne j}^N \Bigg[
      \orb^{*}_i(\vec{r}_1)\,\orb^{*}_j(\vec{r}_2)\,\orb_i(\vec{r}_1)\,\orb_j(\vec{r}_2) \;
        \nonumber \\
&& - \orb^{*}_i(\vec{r}_1)\,\orb^{*}_j(\vec{r}_2)\,\orb_j(\vec{r}_1)\,\orb_i(\vec{r}_2) \; 
      \delta_{\sigma_i,\sigma_j}
\Bigg] \label{eq:slater:n2B_FD_1}\\
&=&\frac{1}{2}
  n(\vec{r}_1)\, n(\vec{r}_2) \nonumber \\
 && - \frac{1}{2}\sum_{ij}^N \Bigg[ \orb^{*}_i(\vec{r}_1)\,\orb^{*}_j(\vec{r}_2)\,\orb_j(\vec{r}_1)\,\orb_i(\vec{r}_2) \; 
      \delta_{\sigma_i,\sigma_j} \Bigg] \\
 &=&  \frac{n(\vec{r}_1)\, n(\vec{r}_2)}{2}  + \frac{J_s(\vec{r}_1,\vec{r}_2)}{2}
 \label{eq:slater:n2B_FD}
\end{eqnarray}
where the exchange term has contributions from likewise spins only. In this form, the sum runs over all $i$ and $j$.
Since the potential from the Hartree contribution is trivial, in the following we concentrate on the functional derivative of the exchange term, $J_s(\vec{r}_1,\vec{r}_2)$.

Applying the product rule of  functional differentiation  yields the expression
\begin{eqnarray}
&& \hspace{-1cm}
\frac{\delta\,J_s(\vec{r}_1,\vec{r}_2)}{\delta n^{\sigma}(\vrs)}   \nonumber \\ 
&=& 
-\sum_{ij}^N \Bigg[ \; \delta_{\sigma_i,\sigma_j,\sigma} \Bigg\{ \nonumber \\ && 
         + \frac{\delta\,\orb^{*}_i(\vec{r}_1)}{\delta n^{\sigma}(\vrs)} \, \orb^{*}_j(\vec{r}_2) \, \orb_j(\vec{r}_1) \, \orb_i(\vec{r}_2) \nonumber \\ &&
         + \orb^{*}_i(\vec{r}_1) \, \frac{\delta\,\orb^{*}_j(\vec{r}_2)}{\delta n^{\sigma}(\vrs)} \, \orb_j(\vec{r}_1) \, \orb_i(\vec{r}_2) \nonumber \\ && 
         + \orb^{*}_i(\vec{r}_1) \, \orb^{*}_j(\vec{r}_2) \, \frac{\delta\,\orb_j(\vec{r}_1)}{\delta n^{\sigma}(\vrs)} \, \orb_i(\vec{r}_2) \nonumber \\ && 
         + \orb^{*}_i(\vec{r}_1) \, \orb^{*}_j(\vec{r}_2) \, \orb_j(\vec{r}_1) \, \frac{\delta\,\orb_i(\vec{r}_2)}{\delta n^{\sigma}(\vrs)} 
 \Bigg\}
 \Bigg] .
 \label{ZM:fderJ}
\end{eqnarray}
In general, the integrals obtained combining~(\ref{ZM:eq:potential_u}), (\ref{eq:slater:n2B_FD}) and (\ref{ZM:fderJ}) 
have the form ($U$ is symmetric in $\vec{r}_1$ and $\vec{r}_2$),
\begin{equation}
 \int \int {\rm d}\vec{r}_1\,{\rm d}\vec{r}_2\,U(\vec{r}_1,\vec{r}_2) \,
     \orb_a^{*}(\vec{r}_2) \,
     \orb_b^{*}(\vec{r}_1) \, 
     \orb_c    (\vec{r}_2) \,
 \frac{\delta\,\orb_d(\vec{r}_1)}{\delta n(\vrs)},
\end{equation}
where $a,b,c,d\in\{i,j\}$. We define the integral $I^{ij}$:
\begin{equation}
 I^{ij}(\vec{r}_1):= \int {\rm d}\vec{r}_2 \, U(\vec{r}_1,\vec{r}_2) \, 
  \orb_i^{*}(\vec{r}_2) \,
  \orb_j    (\vec{r}_2),
 \label{ZM:eq:def:Iij}
\end{equation}
that is known to be smooth. 
The detailed evaluation of these integrals for three-dimensional systems is given in \appendixname~\ref{appendix:Iij}. 

Since $U$ is symmetric in the spatial coordinates  and $I^{ij}=(I^{ji})^*$ equation~(\ref{ZM:fderJ}) can be further simplified. 
We use the fact that 
$\vec{r}_1$ and $\vec{r}_2$ as well as the summation indices $i$ and $j$ can be interchanged, to obtain the expression,
\begin{eqnarray}
v_x^{\sigma}(\vec{r}^\prime) 
&=& \frac{1}{2} \int\int {\rm d}\vec{r}_1 \,  {\rm d}\vec{r}_1 \, U(\vec{r}_1,\vec{r}_2) \frac{\delta J_s(\vec{r}_1,\vec{r}_2)}{\delta n^{\sigma}(\vec{r}^\prime)} \nonumber  \\
&=& - \sum_{ij}^N \Bigg[ \; \delta_{\sigma_i,\sigma_j,\sigma} \Bigg\{ 
            \int {\rm d}\vec{r}\, I^{ji}(\vec{r})\, \orb_j(\vec{r}) \, \frac{\delta \orb_i^{*}(\vec{r})}{\delta n^{\sigma}(\vec{r}^\prime)} \nonumber \\
   &&       +  \int {\rm d}\vec{r}\, I^{ij}(\vec{r})\, \orb_j^{*}(\vec{r}) \, \frac{\delta \orb_i(\vec{r})}{\delta n^{\sigma}(\vec{r}^\prime)}
       \Bigg\}   \Bigg],
\end{eqnarray}
or
\begin{eqnarray}
 \hspace{-0.5cm} v_x^{\sigma}(\vec{r}^\prime) %
&=& - 2 \Re \sum_{ij}^N \delta_{\sigma_i,\sigma_j,\sigma}
         \int {\rm d}\vec{r} \,I^{ij}(\vec{r})\, \orb_j^{*}(\vec{r}) \, \frac{\delta \orb_i(\vec{r})}{\delta n^{\sigma}(\vec{r}^\prime)},
 \label{eq:ZM:exchangePotential}
\end{eqnarray}
where $\delta_{\sigma_i,\sigma_j,\sigma}$ equals $1$ when $\sigma_i=\sigma_j=\sigma$ and vanishes otherwise. 
The exchange potential is always real.
It becomes clear that the crucial quantity of interest is the derivative of an orbital with respect to the density, $\frac{\delta f(\vec{r})}{\delta n({\bf r}^\prime)}$.

\subsection{Explicit Derivatives of Orbitals}
\label{sec:approx_derivative}
Within SIF, the derivative of an orbital with respect to the density is straightforward. Using 
the expansion~(\ref{ZM:orbitalExpansion}) and the definition of the equidensity basis~(\ref{ZM:basis}) we obtain the expression,
\begin{eqnarray}
 \frac{\delta \orb_i^{\sigma_i}(\vec{r})}{\delta n^{\sigma}(\vec{r^{\prime}})}
&=& \sum_{\bf k}  \frac{\delta} {\delta n^{\sigma}(\vec{r^{\prime}})}  a_{\bf k}^{i,\sigma_i}\,    \phi_{\bf k}^{\sigma_i}({\bf r},[n^{\sigma_i}])   \\
&=& \sum_{\bf k}  \Bigg[
\frac{1}{\sqrt{N^{\sigma_i}}}   a_{\bf k}^{i,\sigma_i}\;  e^{i\,\bf k \cdot R}  \;  
    \frac{\delta \sqrt{n^{\sigma_i}(\vec{r})} }{\delta n^{\sigma}(\vec{r^{\prime}})}  \nonumber \\
&&+\sqrt{n^{\sigma_i}(\vec{r}) } \,  a_{\vec{k}}^{i,\sigma_i}\;  e^{i\,\vec{k} \cdot R}  \;  
    \frac{\delta \frac{1}{\sqrt{N^{\sigma_i}}} }{\delta n^{\sigma}(\vec{r^{\prime}})}  \nonumber \\
&& + \sqrt{ \frac{n^{\sigma_i}(\vec{r}) }{N^{\sigma_i}}} \, i \,  a_{\vec{ k}}^{i,\sigma_i}\; e^{i\,\vec{k} \cdot R} \;
  \vec{k} \cdot \frac{\delta R(\vec{r},[n^{\sigma_i}])}{\delta n^{\sigma}(\vec{r^{\prime}})} \nonumber \\
 && + \frac{\delta a_{\bf k}^{i,\sigma_i}}{\delta n^{\sigma}(\vec{r^{\prime}})}
       \, \phi_{\bf k}^{\sigma_i}({\bf r},[n^{\sigma_i}])
\Bigg] \nonumber%
\end{eqnarray}
\begin{eqnarray}
 \frac{\delta \orb_i^{\sigma_i}(\vec{r})}{\delta n^{\sigma}(\vec{r^{\prime}})}
&=&  \sum_{\bf k}  \Bigg[\delta_{\sigma,\sigma_i}
\frac{\delta(\vec{r}-\vec{r}^\prime)}{2 n^{\sigma_i}(\vec{r})} a_{\vec{k}}^{i,\sigma_i} \phi_{\vec{k}}(\vec{r}) \nonumber \\
&& + \delta_{\sigma,\sigma_i}  \sqrt{n^{\sigma_i}(\vec{r}) } \,  a_{\vec{k}}^{i,\sigma_i}\;  e^{i\,\vec{k} \cdot R}  \; 
    \left(  - \frac{1}{2 (N^{\sigma_i})^{3/2}}  \right) \nonumber \\
&&+i \, a_{\vec{k}}^{i,\sigma_i}   \phi_{\vec{k}}^{\sigma_i}({\vec{r}},[n^{\sigma_i}]) \,
  \vec{k} \cdot \frac{\delta R(\vec{r},[n^{\sigma_i}])}{\delta n^{\sigma}(\vec{r^{\prime}})}\nonumber \\
 && + \frac{\delta a_{\bf k}^{i,\sigma_i}}{\delta n^{\sigma}(\vec{r^{\prime}})}
       \, \phi_{\bf k}^{\sigma_i}({\bf r},[n^{\sigma_i}])
\Bigg] \nonumber %
\end{eqnarray}
\begin{eqnarray}
 \frac{\delta \orb_i^{\sigma_i}(\vec{r})}{\delta n^{\sigma}(\vec{r^{\prime}})}
&=&  \delta_{\sigma,\sigma_i} \frac{\delta(\vec{r}-\vec{r}^\prime)}{2 n^{\sigma_i}(\vec{r})} \orb_i^{\sigma_i} (\vec{r}) 
-  \delta_{\sigma,\sigma_i}  \frac{\orb_i^{\sigma_i}}{2\, N^{\sigma_i}} \nonumber \\
 &&+\sum_{\vec{k}} \Bigg[ %
  i \, a_{\vec{k}}^{i,\sigma_i}   \phi_{\vec{k}}^{\sigma_i}({\vec{r}},[n^{\sigma_i}]) \,\;
  \vec{k} \cdot \frac{\delta R(\vec{r},[n^{\sigma_i}])}{\delta n^{\sigma}(\vec{r^{\prime}})} \nonumber \\
 && \hspace{1cm}+ \frac{\delta a_{\bf k}^{i,\sigma_i}}{\delta n^{\sigma}(\vec{r^{\prime}})}
       \, \phi_{\bf k}^{\sigma_i}({\bf r},[n^{\sigma_i}])
\Bigg]. %
\label{eq:dfdn_long}
\end{eqnarray}
For the sake of formal completeness, we include the functional derivative of the normalization integrals $N^\sigma$ with respect to the density, that leads to the second term after the last equals sign.

When the spin of the orbital and the spin of the density with respect to which it is  differentiated coincide, the last expression reduces to the form, 
\begin{eqnarray}
\frac{\delta \orb_i^{\sigma}(\vec{r})}{\delta n^{\sigma}(\vec{r^{\prime}})} 
&=& \frac{\delta(\vec{r}-\vec{r}^\prime)}{2 n^{\sigma}(\vec{r})} \orb_i^{\sigma}(\vec{r}) 
-  \frac{\orb_i^{\sigma}(\vec{r})}{2\, N^{\sigma}} \nonumber \\
&&+ \frac{\delta \vec{R}(\vec{r},[n^{\sigma}])}{\delta n^\sigma(\vec{r}^\prime)} \cdot 
\sum_{\vec{ k}} i \,\vec{k}\, a_{\vec{k}}^{i,\sigma}\;   \phi_{\vec{ k}}({\vec{r}},[n^{\sigma}])
\nonumber\\ &&
 +\sum_{\vec{ k}} \frac{\delta a_{\bf k}^{i,\sigma}}{\delta n^{\sigma}(\vec{r^{\prime}})}
       \, \phi_{\bf k}^{\sigma}({\bf r},[n^{\sigma}]) ,
\label{ZM:FDerivativeOrbital}
\end{eqnarray}
where the second term (from the derivative of the normalization) leads to a constant shift in the potential and can be neglected. 

Clearly,  Eq.~(\ref{ZM:FDerivativeOrbital})  has contributions to the full functional derivative of two distinctly different origins. The third term on the right side contains only terms where the dependence on the density is made explicit by use of the equidensity basis. On the other hand the final term, involving the functional derivative of the expansion coefficients of equidensity basis themselves, has only an implicit dependence on the density. In the following we shall evaluate  Eq.~(\ref{ZM:FDerivativeOrbital}) under the ansatz that this implicit dependence can be neglected - the explicit equidensity basis (EEB) ansatz referred to in the introduction.
While the algebraic consequences of making this ansatz are the subject of ongoing research, the great practical benefit is that the equations that result are of closed form and also become
surprising simple to implement computationally. In addition, results presented for atoms in section~\ref{sec::atoms} indicate that this ansatz produces exchange only total energies of 
similar quality to those of HF and OEP and, furthermore, satisfy expected variational bounds. We will comment on this in more detail in section~\ref{sec:conclusions} while here proceeding with further analysis of the remaining terms in Eq.~(\ref{ZM:FDerivativeOrbital}).

The third term of Eq.~(\ref{ZM:FDerivativeOrbital}) involves an infinite sum over $\vec{ k}$, raising questions about convergence. This problem is circumvented through the realization that the same infinite sum occurs in the gradients of the orbitals.
As shown in \appendixname~\ref{appendix:spacial_trick}, comparing functional and spatial derivatives allows the replacement of the sum over ${\bf  k}$ with expressions involving partial derivatives of the orbitals. 
For example, in the one-dimensional case we obtain the result,
\begin{eqnarray}
 \sum_{k} a^\sigma_k \,i\,k\, \phi^\sigma_k(x)
&=& \frac{N^\sigma}{2\pi\,n^\sigma(x)} \left[\orb^{\prime} -\frac{1}{2\,n^\sigma(x)}  \,\orb(x)\, n^{\sigma \prime}\right] \nonumber \\
&=& \frac{N^\sigma}{2\pi} \frac{1}{\sqrt{n^\sigma(x)}} \left( \frac{f(x)}{\sqrt{n^\sigma(x)}}\right)^\prime ,  \label{ZM:1D:spatialTrick}
\end{eqnarray}
where primes on functions denote spatial derivatives. The three-dimensional case follows analogously, leading to the expression, 
\begin{eqnarray}
&& \sum_{\vk} a_{\vk}^\sigma \phi^\sigma_{\bf k}(\vec{r}) (i \vk)
=:\left( \begin{array}{l} Q_x^\sigma \\Q_y^\sigma \\Q_z^\sigma \end{array} \right) =\vec{Q}^\sigma = \hspace{0cm} \nonumber \\
&& \left( \begin{array}{ll}
\frac{N^\sigma(y,z)}{2\pi n^\sigma(\vec{r})} &\hspace{-6.1pt}\bigg[ 
   \nabla_x\orb-\frac{\orb(\vec{r})}{2n^\sigma(\vec{r})} \frac{\partial n^\sigma(\vec{r})}{\partial x}
   \bigg]  \\[.3cm]
\frac{N^\sigma(z)}{2\pi N^\sigma(y,z)}& \hspace{-6.1pt}\bigg[  \nabla_y\orb 
        - \frac{\orb(\vec{r})}{2n^\sigma(\vec{r})} \frac{\partial n^\sigma(\vec{r})}{\partial y}
        - Q_x^\sigma \beta_{12}^\sigma
     \bigg]  \\[.3cm]
  \frac{N^\sigma}{2\pi\,N^\sigma(z)} &\hspace{-6.1pt}\bigg[
     \nabla_z\orb 
     -\frac{\orb(\vec{r})}{2n^\sigma(\vec{r})} \frac{\partial n^\sigma(\vec{r})}{\partial z}
     - Q_x^\sigma \beta_{13}^\sigma
     - Q_y^\sigma \beta_{23}^\sigma
    \bigg]\hspace{-4.1pt}
 \end{array} \right) , \nonumber\\ &&\label{ZM:def:QQQ}
\end{eqnarray}
with $\beta_{ab}$ being the partial  derivatives of $\vec{R}$, defined in Eq.~(\ref{ZM:R123}), with respect to the coordinates,
\begin{equation}
 \beta_{ab}:=\frac{\partial R_{a}}{\partial b} \qquad \qquad {a,b}\in\{x,y,z\},
\end{equation}
given in \appendixname~\ref{ZM:Appendix:fDerivativeSpacial}.

In short, we can write
\begin{eqnarray}
\frac{\delta \orb_i^\sigma(\vec{r})}{\delta n^{\sigma}(\vec{r^{\prime}})} 
&\simeq& \frac{\delta(\vec{r}-\vec{r}^\prime)}{2 n^{\sigma}(\vec{r})} \orb_i^{\sigma}(\vec{r}) 
-  \frac{\orb_i^{\sigma}}{2\, N^{\sigma}}  \nonumber \\
&&
 + \frac{\delta \vec{R}(\vec{r},[n^{\sigma}])}{\delta n(\vec{r}^\prime)} \cdot \vec{Q}^\sigma,
\label{ZM:FDerivativeOrbitalQ}
\end{eqnarray}
where the $\vec{Q}$ include summations over $\vec{k}$  to infinite order.

Still to be considered is the functional derivative of $\vec{R}$ with respect to the density. The derivation is shown in detail in \appendixname~\ref{ZM:Appendix:fDerivativeR}.  
For the one-dimensional case we obtain the expression, 
\begin{equation}
\frac{\delta R_1(x,[n^{\sigma}])}{\delta n^{\sigma}(x^\prime)}=\frac{2\pi}{N^{\sigma}} \Theta(x-x^{\prime})
\end{equation}
while in three-dimensions we find
\begin{eqnarray}
\frac{\delta R_1(x,y,z,[n^{\sigma}])}{\delta n^{\sigma}(\vec{r}^{\prime\prime})} %
&=& \frac{\delta(y-y^{\prime\prime}) \, \delta(z-z^{\prime\prime})}{N^{\sigma}(y,z)} \times \nonumber \\
   &&
   \Bigg[ 2\pi \, \Theta(x-x^{\prime\prime}) - R_1^\sigma(x,y,z) \Bigg] \label{ZM:dR1dn}\nonumber \\ 
\frac{\delta R_2(y,z,[n^{\sigma}])}{\delta n^{\sigma}(\vec{r}^{\prime\prime})} 
&=& \frac{ \delta(z-z^{\prime\prime})}{N^{\sigma}(z)} \times \nonumber \\
 && 
\Bigg[ 2\pi\, \Theta(y-y^{\prime\prime}) - R_2^\sigma(y,z) \Bigg]\label{ZM:dR2dn} \nonumber \\ 
\frac{\delta R_3(z,[n^{\sigma}])}{\delta n^{\sigma}(\vec{r}^{\prime\prime})} 
 &=& \frac{1}{N^{\sigma}} \Bigg[ 2\pi\,\Theta(z-z^{\prime\prime}) -R_3^\sigma(z)
 \Bigg]  . \label{ZM:dR3dn}
\end{eqnarray}
At this point we derived an analytic and closed-form expression for the functional derivative of an orbital with respect to the density, subject to the conditions imposed by the EEB ansatz described above. 
Through the use of the product rule, these expressions can be used to obtain the contribution, $v_x(\vec{r})$, to the Coulomb potential. 
Computational details and the final expressions for the three-dimensional case are given in \appendixname~\ref{appendix:vx_details}.

\subsection{Formal Summary}
We summarize the discussion of the previous section. 
The use of spatial derivatives allows us to eliminate the explicit evaluation of the equidensity basis and the expansion coefficients, leading to a closed-form expression for the functional derivative of an orbital with respect to the density.
In the one-dimensional case we obtain the expression, 
\begin{eqnarray}
\frac{\delta \orb^\sigma(x)}{\delta n^\sigma(x^{\prime\prime})}
&\simeq& \frac{\orb^\sigma(x) }{2 n^\sigma(x)}  \delta(x-x^{\prime\prime}) - \frac{\orb^\sigma(x)}{2N^\sigma} \nonumber\\ 
&&\hspace{-1.6cm}+  \left(  \frac{1}{n^\sigma(x)}   \left[  \orb^{\sigma\prime}(x)  -\frac{\orb^\sigma(x) \,n^{\sigma\prime}(x)}{2 n^\sigma(x)}   \right]   \right)  \Theta(x-x^{\prime\prime}).
\end{eqnarray}

The analogous results in three-dimensions take the form,  %
\begin{eqnarray}
 \frac{\delta \orb^\sigma(\vec{r})}{\delta n^\sigma(\vec{r}^{\prime\prime})}
&\simeq& \frac{\orb^\sigma(\vec{r}) }{2 n^\sigma(\vec{r})}  \delta(\vec{r}-\vec{r}^{\prime\prime}) - \frac{\orb^\sigma(\vec{r})}{2N^\sigma} \\ \nonumber
&&+ \frac{Q_x^{\sigma}(\vec{r})}{N^\sigma(y,z)} \big[ 2\pi \, \Theta(x-x^{\prime\prime}) -R_1^\sigma(x,y,z) \big] \nonumber \\
    && \hspace{2cm}\delta(y-y^{\prime\prime}) \, \delta(z-z^{\prime\prime}) \nonumber \\
&&+ \frac{Q_y^{\sigma}(\vec{r})}{N^\sigma(z)} \big[ 2\pi \, \Theta(y-y^{\prime\prime}) -R_2^\sigma(y,z) \big] \nonumber \\
    && \hspace{2cm}\delta(z-z^{\prime\prime}) \nonumber \\
&&+ \frac{Q_z^{\sigma}(\vec{r})}{N^\sigma} \big[ 2\pi \, \Theta(z-z^{\prime\prime}) -R_3^\sigma(z) \big] ,
\end{eqnarray}
with the quantities $R$ defined in (\ref{ZM:R123}), the $N$ in (\ref{ZM:def:NNN}) and the $\vec{Q}$ in (\ref{ZM:def:QQQ}).
Computational details are given in \appendixname~\ref{appendix:vx_details}. The calculation of the potential becomes now straightforward using equations~(\ref{eq:ZM:exchangePotential}) and (\ref{ZM:eq:def:Iij}).

\subsection{Iterative Procedure}
The solution of the \KS equations relies on an iterative procedure, 
with an updated \KS potential and density determined at each iteration step. 
The sequence of steps in the treatment of the \KS equations is summarized below:

\begin{enumerate}
\item{At each step, $i$, of the iteration, determine the orbitals, $f_j^{(i)}({\bf r})$, 
and the density, $n^{(i)}({\bf r})$.}
\item{Determine the derivatives of the orbitals with the respect to the density, 
take the functional derivative of the Coulomb energy with respect to the density in terms of spatial gradients, and 
obtain the Coulomb energy contribution to the \KS potential. }
\item{Solve the \KS equation for the new potential, go back to the first step  and iterate until convergence is reached within some preset tolerance.}
\end{enumerate}
In short, the only difference with conventional procedures is the treatment  of the full Coulomb potential expressed in terms of the pair density, rather than just the Hartree term or modifications to it. 
\section{Examples}
\label{sec:examples}
In this section we present the results of applications of our method to
two systems for which the exact answers are known analytically, to a one dimensional model system, and finally on the series of atomic systems from Helium to Krypton.

\subsection{Analytic Examples}
The formalism allows the expression of the functional derivatives of the exchange energy 
with respect to the density by analytic means and leads to closed-form expressions in terms of 
the spatial gradients of the orbital functions.
These expressions  can be compared to exact results in cases  where analytic expressions are known.

\subsubsection{Two-Electron Systems}
As a first example we discuss the ground state of a two-electron systems, such as the Helium atom or the Hydrogen dimer ($H_2$), 
with two electrons of opposite spin in the same spatial orbital under the same external potential. 

Assuming  real, and node free orbitals we obtain
\begin{eqnarray}
 n^\sigma(\vec{r})&=&\sum_{i=1}^1 \left|\orb_i(\vec{r}) \right|^2 =  \left|\orb(\vec{r}) \right|^2 \\
 \orb(\vec{r})&=&\sqrt{n^\sigma(\vec{r})}\\
 \frac{\delta \orb (\vec{r}) }{\delta n^\sigma(\vec{r}^\prime)} &=& \frac{1}{ 2\, \sqrt{n^\sigma(\vec{r})}} \delta(\vec{r}-\vec{r}^\prime)\\
\orb  \frac{\delta \orb }{\delta n^\sigma} &=& \frac{1}{2} \delta(\vec{r}-\vec{r}^\prime),
\end{eqnarray}
from which the exchange potential can be determined as follows:
\begin{eqnarray}
 v_x^{\sigma}(\vec{r}) 
 &=& \frac{1}{2} \int\int {\rm d}\vec{r}_1 \, {\rm d}\vec{r}_2 \; U(\vec{r}_1,\vec{r}_2) 
              \frac{\delta J_s(\vec{r}_1,\vec{r}_2) }{\delta n^\sigma(\vec{r})} \\
 &=& - \int\int {\rm d}\vec{r}_1 \, {\rm d}\vec{r}_2 \; U(\vec{r}_1,\vec{r}_2) \bigg\{\nonumber\\
 &&      \orb(\vec{r}_1) \frac{\delta \orb(\vec{r}_1) }{\delta n^\sigma} \, \orb(\vec{r}_2) \orb(\vec{r}_2)\nonumber\\
 &&   +  \orb(\vec{r}_2) \frac{\delta \orb(\vec{r}_2) }{\delta n^\sigma} \, \orb(\vec{r}_1) \orb(\vec{r}_1)
     \bigg\} \\
&=&   - \int\int {\rm d}\vec{r}_1 \, {\rm d}\vec{r}_2 \; U(\vec{r}_1,\vec{r}_2) \bigg\{\nonumber\\
  &&  \frac{1}{2}\delta(\vec{r}-\vec{r}_1) \, \orb(\vec{r}_2) \orb(\vec{r}_2) \nonumber\\
  && + \frac{1}{2}\delta(\vec{r}-\vec{r}_2) \, \orb(\vec{r}_1) \orb(\vec{r}_1)
      \bigg\} \\
&=& -\int {\rm d}\vec{r}_1  \; U(\vec{r}_1,\vec{r}) n^\sigma(\vec{r}_1)\nonumber\\
v_x^{\sigma}(\vec{r}) &=& -\frac{1}{2}  \int {\rm d}\vec{r}_1 \,  \; U(\vec{r}_1,\vec{r}) \,n(\vec{r}_1) \nonumber\\
&=& 
-\frac{1}{2} v_{\rm H}(\vec{r}).
\end{eqnarray}
The exchange potential is exactly half of the Hartree potential but with the opposite sign, so that the self-interaction error is half of the Hartree term. \\
Using the method proposed in this paper, we write
\begin{equation}
 f(\vec{r})=\sum_{\vec{k}=\vec{0}} a_{\vec{k}} \phi_{\vec{k}}(\vec{r})\quad \mbox{with}\quad \, a_{\vec{0}}=1,
\end{equation}
where all coefficients  other  than $\vec{k}=\vec{0}$ vanish. From Eq.~(\ref{ZM:FDerivativeOrbital})  we obtain the expression,
\begin{eqnarray}
 \frac{\delta f(\vec{r})}{\delta n^{\sigma}(\vec{r}^\prime)}
&=& \frac{\delta(\vec{r}-\vec{r}^\prime)}{2n^\sigma(\vec{r})}\orb(\vec{r}) = \frac{\delta(\vec{r}-\vec{r}^\prime)}{2\orb(\vec{r})}.
\end{eqnarray}
Using Eq.~(\ref{eq:ZM:exchangePotential}) for the exchange potential yields the result,
\begin{eqnarray}
 v_x^\sigma(\vec{r}^\prime) 
&=& -2 \int  {\rm d}\vec{r} \; I^{11} (\vec{r}) \, \orb(\vec{r})\, \frac{\delta \orb(\vec{r})}{\delta n^\sigma(\vec{r}^\prime)} \\
&=& -I^{11}(\vec{r}^\prime)=-\int {\rm d}\vec{r} \; U(\vec{r},\vec{r}^\prime) \orb(\vec{r}) \, \orb(\vec{r}) \\
&=& -\int {\rm d}\vec{r} \; U(\vec{r},\vec{r}^\prime) n^{\sigma}(\vec{r}) \\
v_x^\sigma(\vec{r}^\prime) &=&  -\frac{1}{2} \int {\rm d}\vec{r} \; U(\vec{r},\vec{r}^\prime) n(\vec{r}) %
=-\frac{1}{2} \, v^{\rm{H}}(\vec{r}^\prime).
\end{eqnarray}
For this simple example, the SIF method reproduces the correct analytic expression for the exchange potential.

\subsubsection{Hartree Potential}
Our second example deals with the Hartree term. The functional derivative of the Hartree energy $E_{\rm H}$ with respect to the density 
is well known.  
From 
\begin{equation}
 E_H=\frac{1}{2} \int\int {\rm d}\vec{r} \,{\rm d}\vec{r}^\prime \, \frac{n(\vec{r})\, n(\vec{r}^\prime)}{\left|\vec{r}-\vec{r}^\prime \right|}
\end{equation}
we obtain
\begin{equation}
 \frac{\delta E_{\rm H}}{\delta n(\vec{r}^{\prime\prime})} = v_{\rm H}(\vec{r}^{\prime\prime})
= \int {\rm d}\vec{r} \frac{n(\vec{r})}{\left|\vec{r}-\vec{r}^{\prime\prime} \right|}.
\end{equation}
It can also be shown, that
\begin{equation}
 \frac{\delta E_{\rm H}}{\delta n(\vec{r}^{\prime\prime})}
= \frac{\delta E_{\rm H}}{\delta n^{\uparrow}(\vec{r}^{\prime\prime})}
= \frac{\delta E_{\rm H}}{\delta n^{\downarrow}(\vec{r}^{\prime\prime})}
=v_{\rm H}(\vec{r}^{\prime\prime}).
\end{equation}
We now express the density in terms of orbitals, 
\begin{equation}
 n^\sigma(\vec{r}) = \sum_{i=1}^{N^\sigma} \orb^{\sigma *}_i(\vec{r}) \, \orb_i^\sigma(\vec{r}),
\end{equation}
and express the orbitals in terms of the equidensity basis, and take the functional derivative of $E_{\rm H}$  written in terms of the expanded forms. The results leads to $v_{\rm H}$. 
The detailed derivation of this result is shown in \appendixname~\ref{appendix:VHartree}. 

Notabley, despite the use of our 
equidensity ansatz in  Eq.~(\ref{ZM:FDerivativeOrbital}), we stall obtain the correct analytical result for this case.

\subsection{Numerical Examples}
In this section we apply our method to one-dimensional systems in terms of the particles-in-a-box problem and to realistic atomic systems.

\subsubsection{One-dimensional Square Well}
The work reported in this section is designed to test a simple case of non-interacting, spinless Fermions
confined in an one-dimensional well of length, $L=|x_1-x_0|=1$,  with infinite potential walls. 
We choose $N=6$. 
This example is used for illustrating the method, deriving the potential corresponding to the energy for a given form of the inter-particle interaction.  
We restrict ourselves to a non-self consistent solution for a vanishing (or constant) potential. 
We choose an inter-particle interaction that decays exponentially
with respect to inter-particle distance,
\beq
U(x_1,x_2)=\Lambda \, {\rm e}^{-\lambda|x_1-x_2|},
\label{ExpU.1}
\eeq
that allows us the freedom of  manipulating the range of the interaction and assess its 
effects on the exchange potential.\\

For this case the orbitals corresponding to $N$ lowest energies are known analytically and
lead to an analytic expression for the density in Eq.~(\ref{Slatern.1}) and
the pair density, Eq.~(\ref{eq:slater:n2B_FD}). 
The normalized wave functions of this system are given by the expressions,
\begin{equation}
f_{n}(x)=\sqrt{\frac{2}{L}}  \sin{\left(\frac{n \,\pi\, x}{L}\right)}  \qquad 0\le x \le L,
\label{eq:1D:f}
\end{equation}
with quantum numbers $n=1,2,3,\dots$.
The corresponding energies are
\begin{equation}
E_n=\frac{\hbar^2}{2m} \left(\frac{n\pi}{L}\right)^2,
\end{equation}
with the ground-state density given by $n(x)=\sum_{n=1}^{6}|f_n(x)|^2$.

\begin{figure}[htb]
\usepictures{
\begin{center}
   \includegraphics[width=0.45\textwidth,clip]{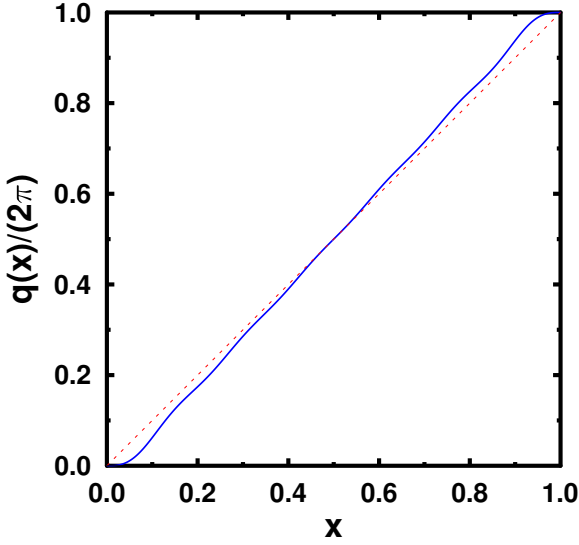}
 \end{center}
}
 \caption{The quantity $q(x)$ (solid blue line) for the one-dimensional square well example
          with 6 electrons. The red dashed line shows $q(x)=x$ which corresponds to a constant density. In this case the basis functions $\phi$ become plane waves.}
 \label{fig:q_of_x}
\end{figure}

\begin{figure}[ht]
\usepictures{
\begin{center}
   \includegraphics[width=0.45\textwidth]{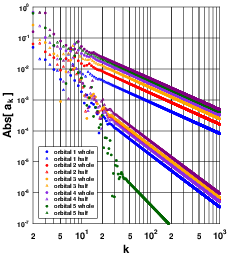}
 \end{center}
}
 \caption{Log-log plot of expansion coefficients of spatially symmetric (even under reflection)
  and antisymmeytric (odd under reflection) orbitals in terms of an equidensity basis, as indicated 
  in the legend.}
 \label{Coeff.1}
\end{figure}

In analogy to Eq.~(\ref{ZM:R123}), we define the quantities $R_{1/2/3}(x,y,z)$, that  in one dimension reduce~\cite{PARR1} to a function $q(x)$,
\begin{equation}
q(x)=\frac{2\,\pi}{N} \int_{x_0}^{x} n(x^\prime)\,{\rm d}x^\prime.
\label{eq:1D:qDef}
\end{equation}
For the six-electron case, $q(x)$ is shown in Fig.~\ref{fig:q_of_x}.
We see that the function $q(x)$ is almost linear. An exact linear behavior, $q(x)=x$, 
(red dashed line) would correspond to a constant density with the equidensity orbitals 
reduced to plane-waves. \\
The functional derivative of $\frac{\delta q(x)}{\delta n(x^{\prime\prime})}$ takes the form,
\beq
\frac{\delta q(x)}{\delta n(x^{\prime\prime})}=\frac{2\pi}{N}\Theta(x-x^{\prime\prime}).
\label{FuncDer.1}
\eeq
We also have, neglecting terms which would contribute only to a constant shift in the potential,
\beqn
\nonumber &&\frac{\delta f_n(x)}{\delta n(x^{\prime\prime})}=\frac{f_n(x)}{2n(x)}\delta(x-x^{\prime\prime})
+\frac{\delta q(x)}{\delta n(x^{\prime\prime})}\sum_k a^n_k{\rm i}k\phi_k[n(x)].
\label{Sumkk.1}
\eeqn
In spite of the dependence on $k$, the sum on the right-hand side converges sufficiently rapidly 
to be numerically stable. In this case, the equidensity basis functions, $\phi_k(x)$, take the form,
\begin{equation}
\phi_k(x)=\sqrt{\frac{n(x)}{N}} \; e^{ i\,k\,q(x)}=\sqrt{\frac{n(x)}{N}} \; e^{ i\,k\,q(x,[n])}.
\end{equation}
There exist two choices for $k$ in constructing a complete and orthonormal set of basis functions: 
Either signed whole integer values, $k=0,\pm 1,\pm 2,\pm 3,\dots$, or half-integer values, 
$k=\pm \frac{1}{2},\pm \frac{3}{2},\pm \frac{5}{2},\dots$. 
The rate of convergence of the expansion of the orbitals is found to depend on spatial symmetry, 
with those even under reflection (symmetric) about the center of the box being described more 
efficiently by whole integer values of $k$, while half-integer values lead to faster convergence 
of the orbitals that are odd under reflection (antisymmetric). 
A plot of the coefficients vs. values of $k$ is shown in Fig. \ref{Coeff.1}.
We find that with the choice of the faster converging expansion the error in the
norm becomes smaller than $10^{-5}$ when more than 100 basis functions are taken into account. 
Generally, we find that fewer than 1,000 functions are sufficient for convergence. 

\begin{figure}[ht]
\usepictures{
\begin{center}
  \subfigure{\includegraphics[width=0.46\textwidth]{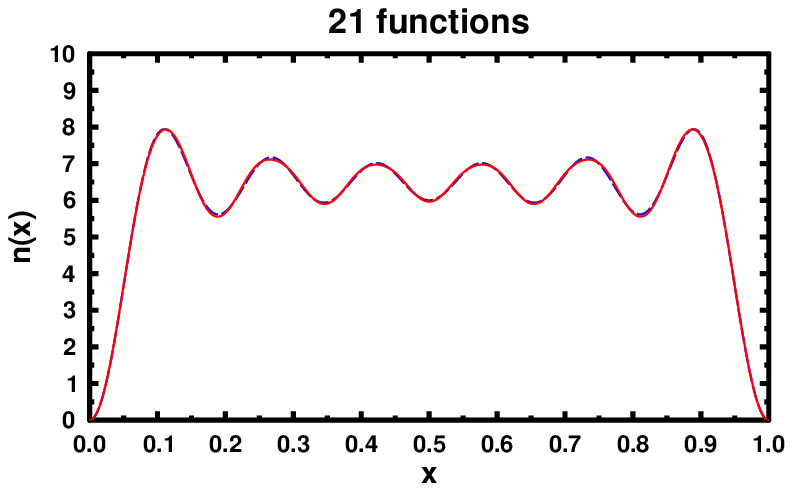}} \\ [-.2cm]
  \subfigure{\includegraphics[width=0.46\textwidth]{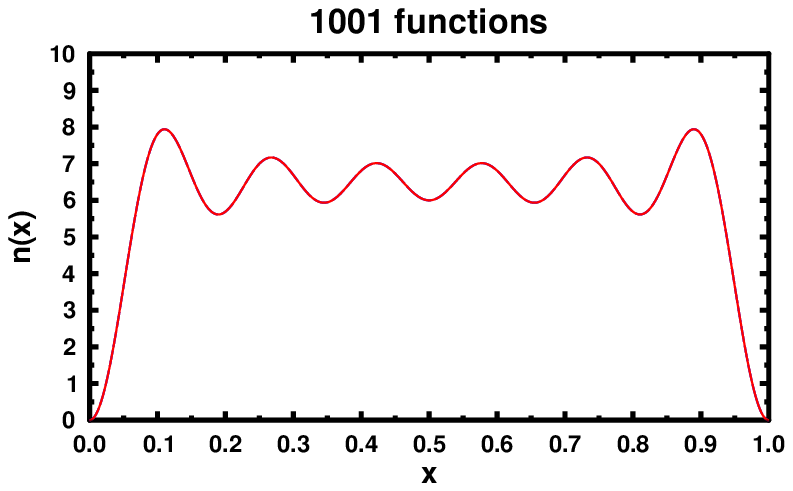}}
 \end{center}
}
 \caption{Ground-state density of six non-interacting electrons in an one-dimensional
          box with infinite walls as discussed in the text. The blue dashed line marks the 
          analytic expression, plotted below the other curve. At this scale they are almost indistinguishable.}
 \label{Density.1}
\end{figure}

\begin{figure}[ht]
\usepictures{
\begin{center}
   \includegraphics[width=0.45\textwidth]{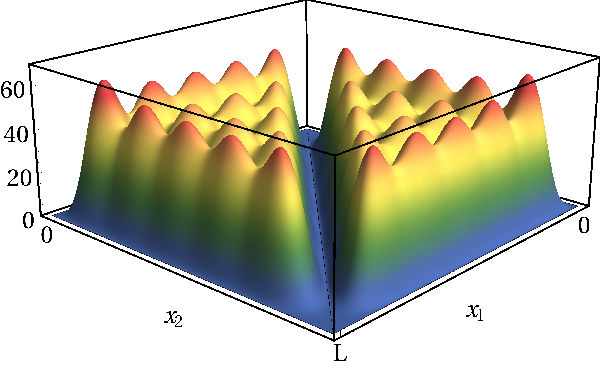}
 \end{center}
}
 \caption{Non-interacting Ground-state pair density of six non-interacting electrons in an one-dimensional
          box with infinite walls as discussed in the text.}
 \label{Pair.1}
\end{figure}

The ground-state density of the system, (the six electrons occupying the orbitals
labeled $n=1,2,\dots,6$), is shown in Fig~\ref{Density.1}, while the pair density
is shown in Fig.~\ref{Pair.1}, with the number of basis functions indicated in the panel. 
Within the resolution of the figure, the two results are essentially indistinguishable. 
The same rate of convergence characterizes the pair density, shown in Fig.~\ref{Pair.1}. We note the vanishing of the 
pair density along the line $x_1=x_2$, as expected from Eq.~(\ref{eq:slater:n2B_FD_1}).

\begin{figure}[ht]
\usepictures{
\begin{center}
  \subfigure{\includegraphics[width=0.43\textwidth]{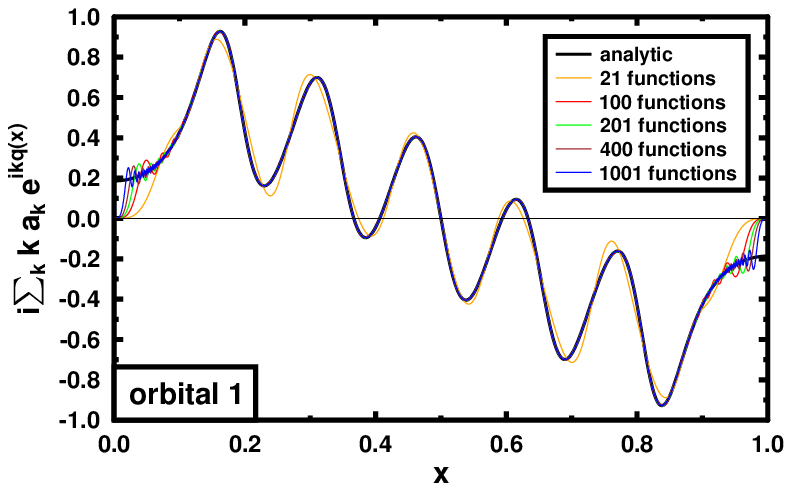}} \\ [-.2cm]
  \subfigure{\includegraphics[width=0.43\textwidth]{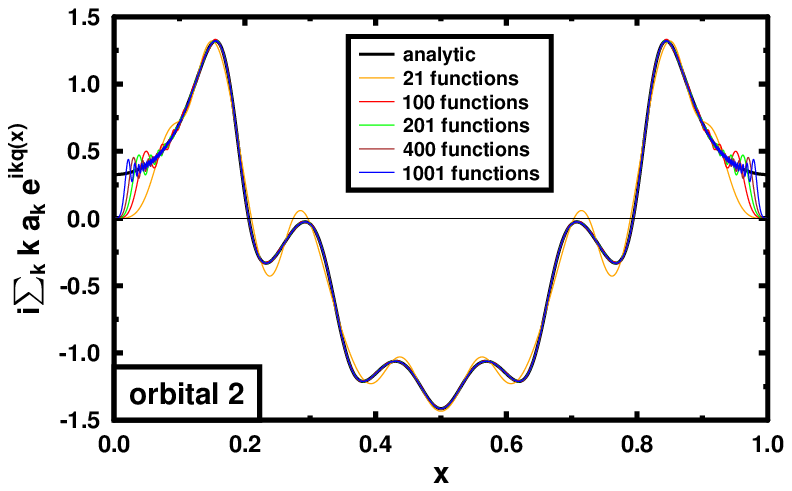}} \\ [-.2cm]
  \subfigure{\includegraphics[width=0.43\textwidth]{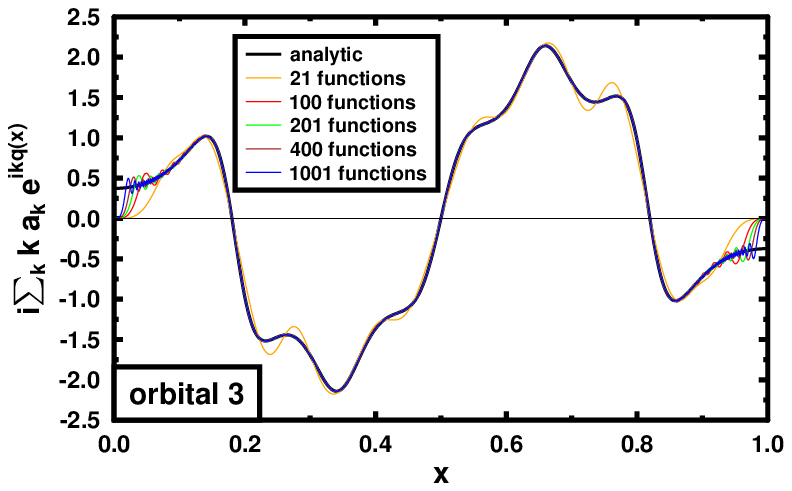}} 

\end{center}
}
 \caption{Infinite sum, see Eq.~(\ref{ZM:1D:spatialTrick}), evaluated for the first three orbitals of the six electron example.  
   Convergence is shown by increasing the numbers of coefficients $a_k$ (chosen symmetrically around $0$) in comparison to the analytic result.}
 \label{fig:infinite_sum}
\end{figure}

Another quantity to look at for convergence is the infinite sum over $k$ appearing in Eq.~(\ref{ZM:FDerivativeOrbital}). 
The result with the sums  carried out to infinite order is given by the analytic expression in Eq.~(\ref{ZM:1D:spatialTrick}).  
The sums over $k$ for the first three orbitals are shown in Fig.~\ref{fig:infinite_sum} for different numbers of expansion coefficients and basis functions 
taken into account, with those used chosen symmetrically around $k=0$. 
For this example, we find a rather good agreement using a few tens of coefficients in a broad range in space, between about $x=0.1$ and $x=0.9$ in Fig.~\ref{fig:infinite_sum}.
At the boundaries, the difference between the analytic expression and the truncated sums increases, the oscillations being a manifestation of the  Gibbs phenomenon.
Even though the results show convergence with the number of terms taken into account, we use the closed-form expressions of Eq.~(\ref{ZM:1D:spatialTrick}). 
This bypasses the explicit construction of the equidensity basis and the expansion coefficients.  

\begin{figure}[ht]
\usepictures{
\begin{center}
  \subfigure{\includegraphics[width=0.23\textwidth]{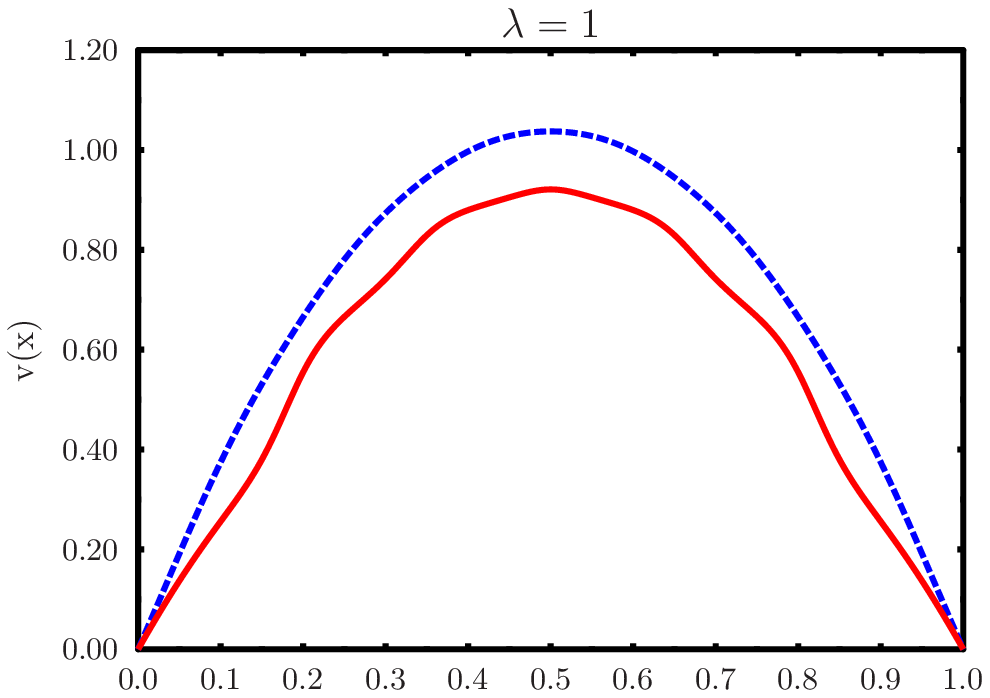}} %
  \subfigure{\includegraphics[width=0.23\textwidth]{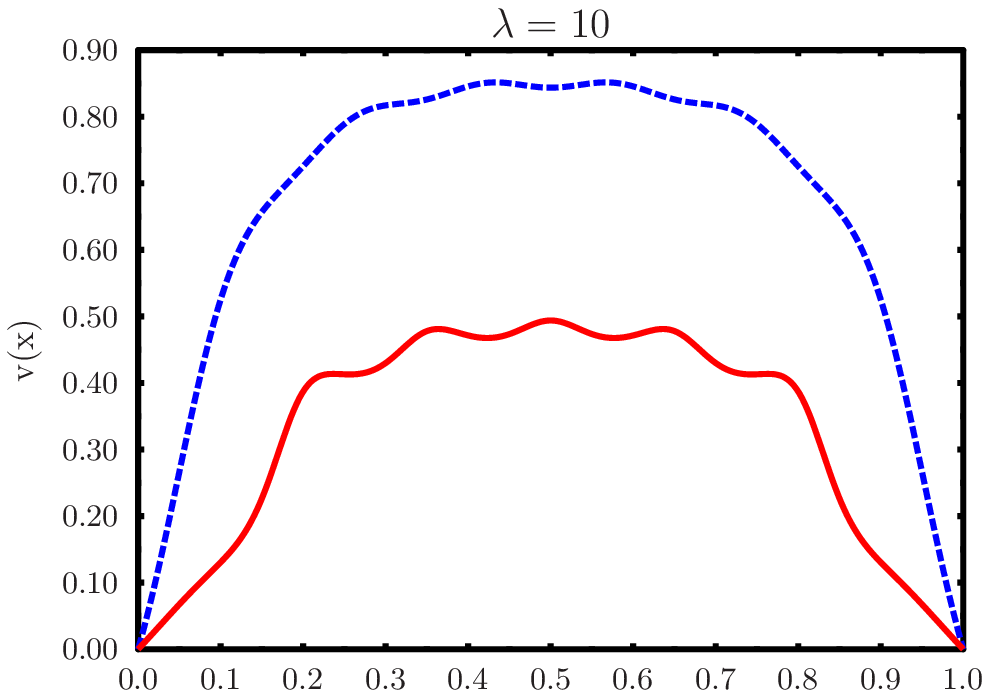}} \\ [-.2cm]
  \subfigure{\includegraphics[width=0.23\textwidth]{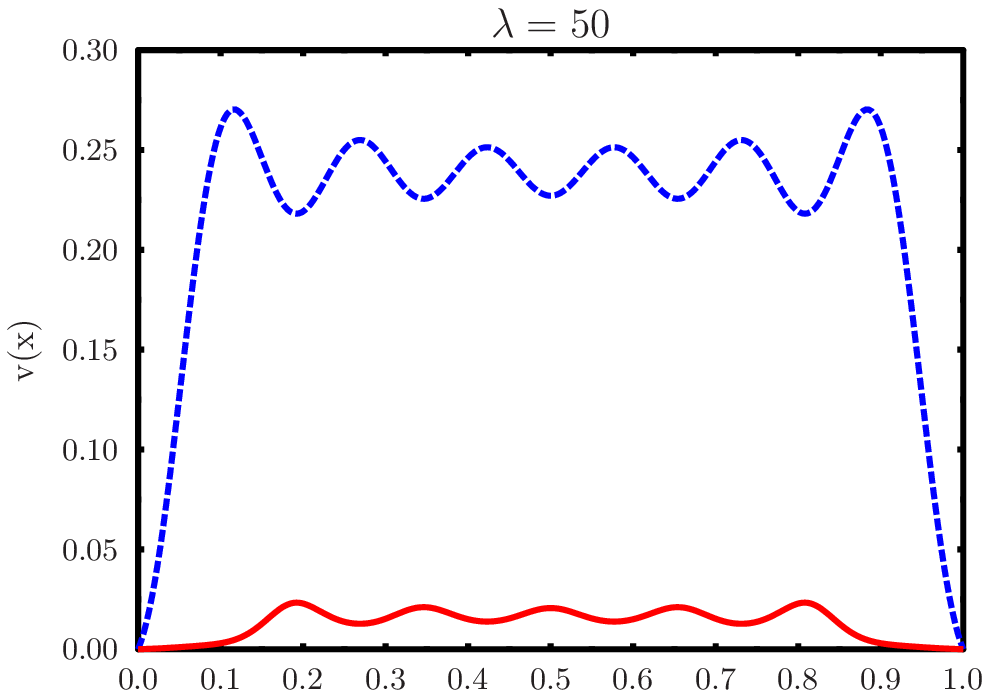}} %
  \subfigure{\includegraphics[width=0.23\textwidth]{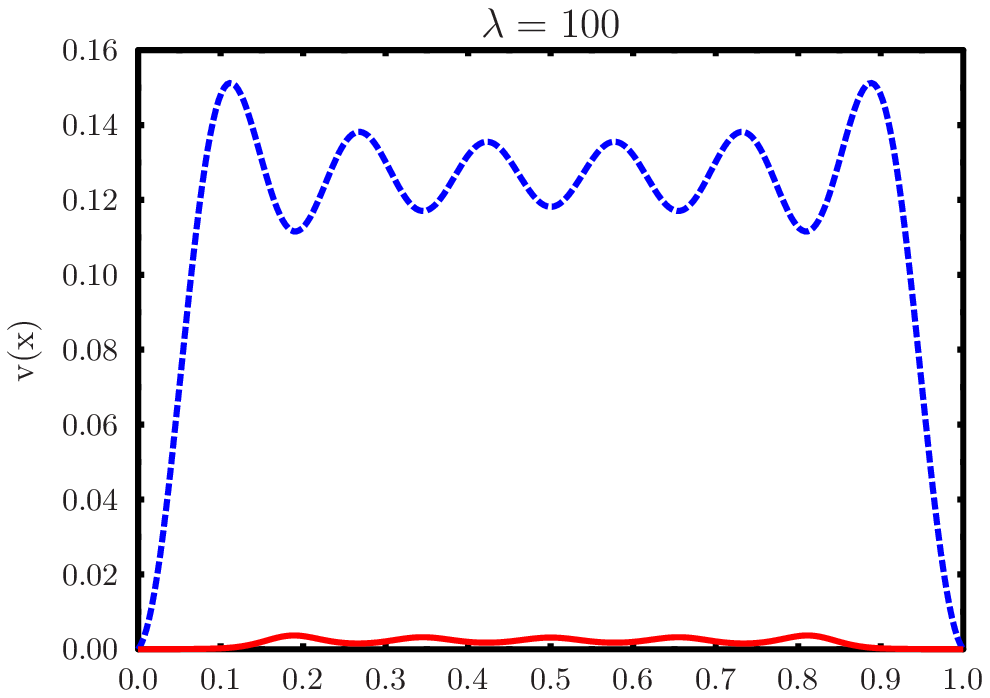}} %
\end{center}
}
 \caption{Single-particle potential contributed by the interaction energy in the system of
          six non-interacting electrons confined in a box with infinite walls as discussed in the text.
          The blue dashed line is calculated from~Eq.~(\ref{Ucl1}) (Hartree), whereas the red solid line is based on Eq.~(\ref{QMU1}) (pair density).}
 \label{Pot.1}
\end{figure}

Fig.~\ref{Pot.1} shows the potential obtained for the four values of $\lambda=1,10,50,100$, 
as defined in (\ref{ExpU.1}),
using both the Hartree expression (blue dashed curves) for the 
interaction energy,
Eq.~(\ref{Ucl1}), as well as the quantum mechanically correct expression,
Eq.~(\ref{QMU1}) (red solid line). In all cases, convergence is well established with about 200 basis functions,
although the results shown here correspond to 1,000 terms taken into account. The analytic expressions lead to the same result. 
An immediate effect of the quantum expression are the lower values of the potential
obtained from Eq.~(\ref{QMU1}), caused by the presence of the exchange term that reduces the
effective region of interaction of two electrons. 
Of interest is also the behavior of the two different expressions for the potential in the
limit of short-range interaction, or large values of $\lambda$.
The Hartree term yields a potential
that approaches the form of the density, essentially reaching that form for $\lambda=100$
(upper curve in panel on lower right-hand corner).
In that limit, the interaction potential resembles a delta-function in two-particle space
and the energy given by the Hartree term is simply proportional to the density at the point $x_1=x_2$.%

The behavior is considerably different when the pair density is used to calculate the
interaction energy.
Now, in the limit of short-range interaction interaction, and spinless fermions, 
two particles cease
interacting altogether because the pair density vanishes in the region in which the interaction
is noticeably non-zero and the contribution to the
single-particle potential drops essentially to zero (lower curve in panel in lower right-hand corner 
in the figure).%

\subsubsection{Atomic Calculations}
\label{sec::atoms}
In the following we apply the SIF method to realistic atomic systems. 
Details of the calculations, in particular of the evaluation of the Coulomb and exchange potential are given in \appendixname~\ref{appendix:vx_details}.
Because the form of the correlation energy is currently unknown, the results presented are all fully self-consistent and converged within the exchange-only mode ($E_c=0$). 
This allows us to compare SIF results with those obtained within the exchange-only OEP and the \HF methods.

\begin{figure*}[ht]
\usepictures{
\begin{center}
   \includegraphics[width=0.90\textwidth]{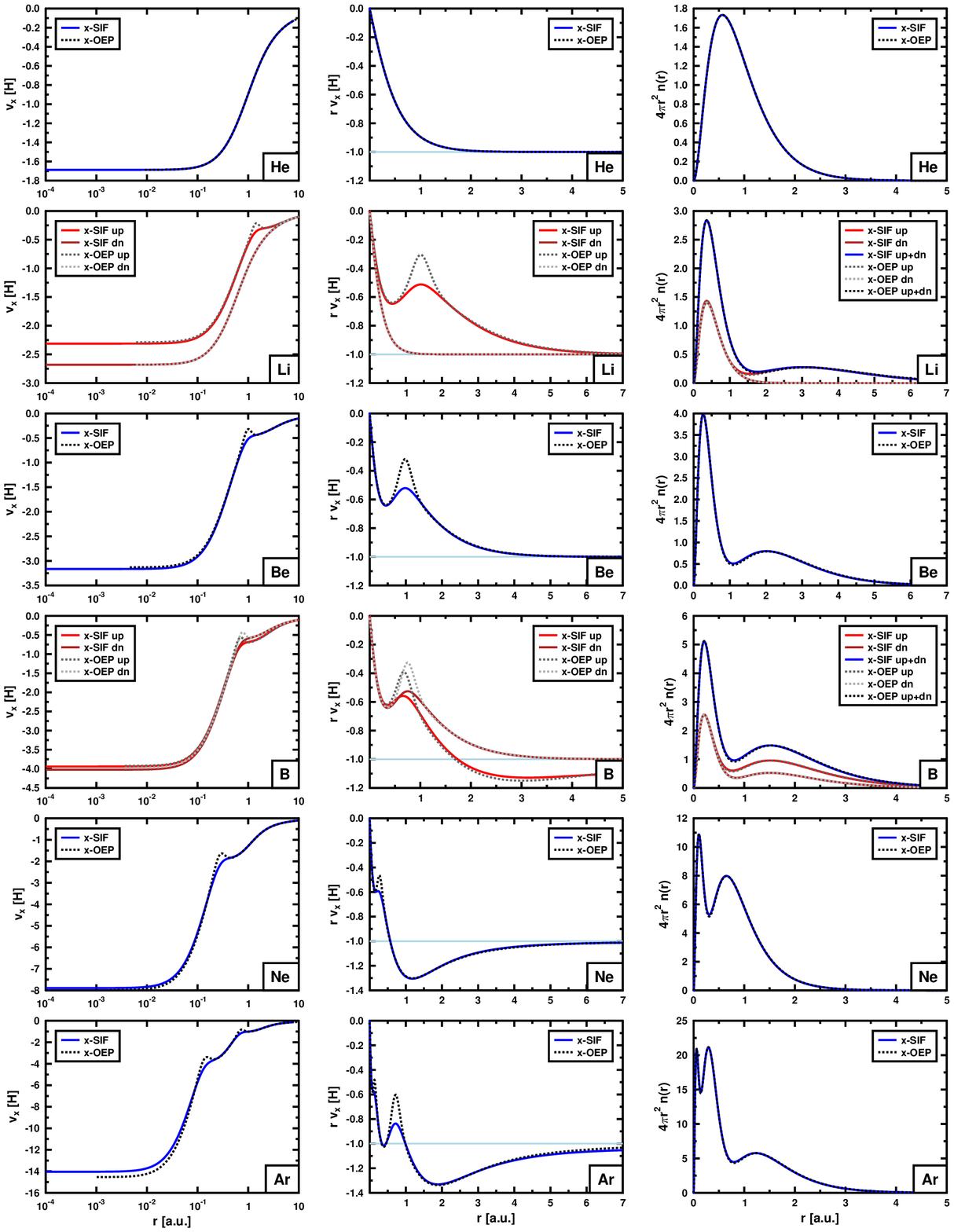}
 \end{center}
}
 \caption{Exchange potentials for Helium, Lithium, Beryllium, Boron, Neon and Argon. The left columns depicts the radial 
   exchange potential on a logarithmic scale, the middle column the exchange potential multiplied by $r$, the distance from the nucleus,  so as to exhibit its 
   asymptotic behavior, and the third column  the density.
   Results of this paper (referred as x-SIF) are compared to OEP calculations. 
  The curves represent self-consistent results obtained in the absence of a correlation energy (exchange only).}
 \label{pic:HeLiBeBNeAr}
\end{figure*}

Fig.~\ref{pic:HeLiBeBNeAr} show the results for Helium, Lithium, Beryllium, Boron, Neon and Argon obtained from a self-consistent exchange-only calculation.
The left column shows the exchange potential on a logarithmic scale that allows a detailed study of the behavior  close to the nucleus. 
To emphasize the asymptotic behavior,  in the middle panel,  the same exchange potential is shown on a linear scale, but multiplied by the distance to the nucleus, $r$. 
Finally, the density is plotted in the right column. 
For all examples, the corresponding full OEP  results are shown for comparison. 
Lithium and Boron are spin-polarized, resulting in different potentials for the majority and minority spin channel.

For Helium, the exchange potential within SIF coincides with the one obtained by the OEP method, and therefore the densities are also identical.
This identity is not preserved for heavier atoms. 
As seen in Fig.~\ref{pic:HeLiBeBNeAr}, the SIF potential has always a non-negative slope, whereas the OEP potentials exhibits  regions where the slope becomes  negative. 
In the OEP literature~\cite{OEP1} this is referred to as the intershell structure. %

\begin{figure}[ht]
\usepictures{
\begin{center}
  \subfigure{\includegraphics[width=0.47\textwidth,clip]{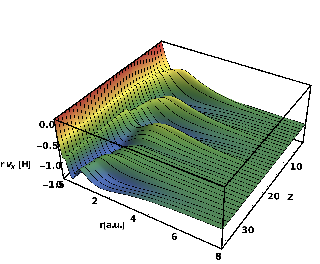}} \\ [-.2cm]
  \subfigure{\includegraphics[width=0.47\textwidth,clip]{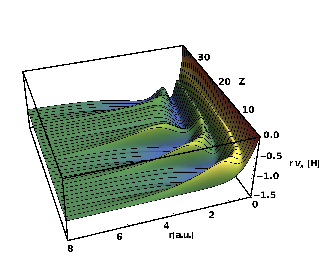}} %
 \end{center}
}
 \caption{SIF exchange potentials multiplied by $r$, the distance from the nucleus, for the atomic series Helium ($Z=2$) to Krypton ($Z=36$). 
 Both figures show the same data, just at a different angle. The colored surface is for illustration, 
  and exchange potentials are only meaningful for integer $Z$ values, marked by the black lines.}
 \label{fig:Vx_atom_series}
\end{figure}

Even though the potentials are different, the SIF and OEP densities are indistinguishable at the scale of the plot.
All exchange potentials behave regularly near the origin, and have an $-1/r$ asymptotic behavior, as expected.
For illustrative purposes, we  plot the exchange potentials for the atoms Helium to Krypton  in Fig.~\ref{fig:Vx_atom_series}.

\begin{figure}[ht]
\usepictures{
\begin{center}
   \includegraphics[width=0.45\textwidth]{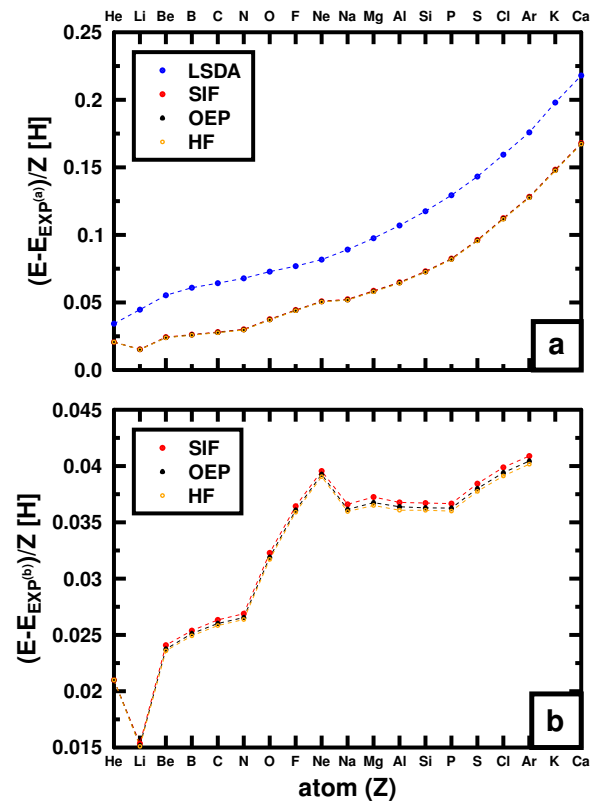}
 \end{center}
}
 \caption{Total energy differences, normalized by the number of electrons,  between SIF, x-OEP, and \HF to the 
   experimental values, see Table~\ref{table::atomic_energies}. a) refers to EXP$^{(a)}$, and b) to the column EXP$^{(b)}$. 
For comparison, LSDA energies~\cite{atoms_LDA_etot} are shown in panel a).}
 \label{pic:atomic_energies}
\end{figure}

\begin{figure}[htp]
\usepictures{
\begin{center}
    \includegraphics[width=0.45\textwidth]{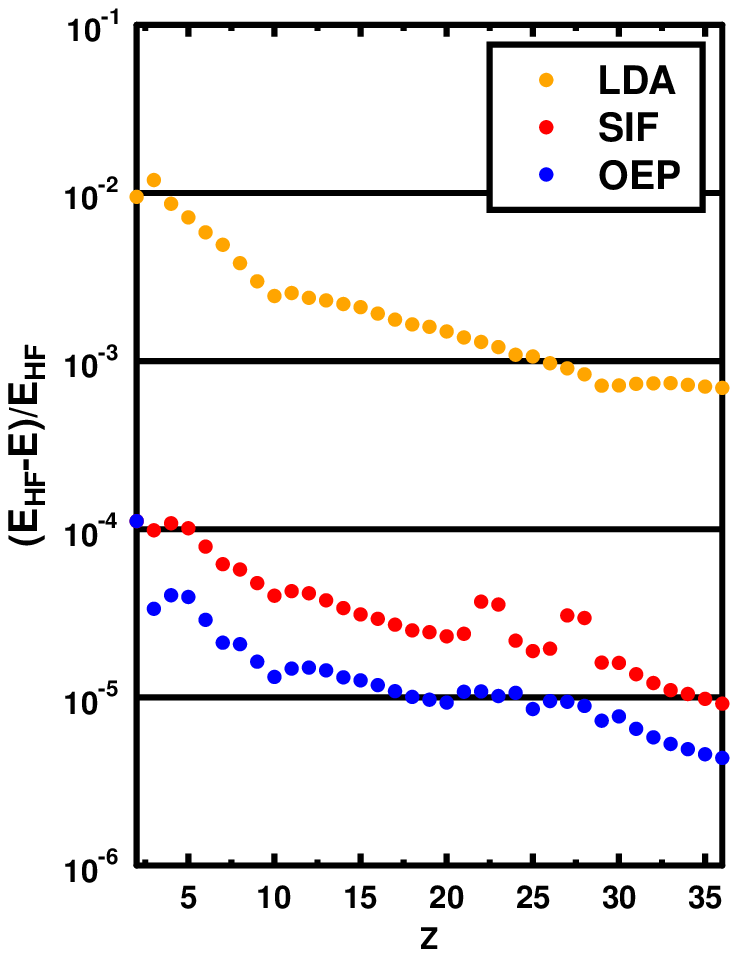}
 \end{center}
}
 \caption{Total energy differences relative to Hartree-Fock of LDA~\cite{atoms_LDA_etot}, SIF and OEP\cite{PhysRevA.47.165} for the atom series from He to Kr, plotted on a logarithmic scale. 
  Note that LDA energies also include the usual LDA correlation energy, where SIF and OEP energies are exchange only. }
 \label{pic:errors}
\end{figure}

Total energies for the atom series obtained in SIF, OEP and HF are compared with experimental determined ones in Table~\ref{table::atomic_energies} and are plotted as difference to experiment in Fig.~\ref{pic:atomic_energies} as absolute differences normalized to the atomic number, ($(E-E_{EXP})/Z$). In Fig.~\ref{pic:errors} we also plot the fractional error $(E-E_{HF})/E_{HF}$, relative to the corresponding HF energies, which gives a clearer picture of the rather small relative "errors", as compared to LDA, of both OEP and SIF
\begingroup
\begin{table*}
\begin{center}
\begin{tabular}{| r | c | l | l | l| r | r |r|}
\hline
 $Z$ & Symbol & Element  & EXP$^{(a)}$  & EXP$^{(b)}$(Ref.\cite{PhysRevA.47.3649})      & x-SIF& x-OEP (Ref.\cite{PhysRevA.47.165})  & HF (Ref.\cite{PhysRevA.47.165}) \\
\hline
  2 & He     & Helium           & -2.90339             & -2.904    & -2.862     &   -2.862  &   -2.862 \\ 
  3 & Li     & Lithium          & -7.47798             & -7.47806  & -7.432     &   -7.433  &   -7.433 \\ 
  4 & Be     & Beryllium        & -14.6685             & -14.66736 & -14.571    &  -14.572  &  -14.573 \\ 
  5 & B      & Boron            & -24.6582             & -24.65391 & -24.527    &  -24.528  &  -24.529 \\ 
  6 & C      & Carbon           & -37.8557$\pm$0.0002  & -37.8450  & -37.687    &  -37.689  &  -37.690 \\ 
  7 & N      & Nitrogen         & -54.6119             & -54.5892  & -54.401    &  -54.403  &  -54.405 \\ 
  8 & O      & Oxygen           & -75.1100$\pm$0.0001  & -75.0673  & -74.809    &  -74.812  &  -74.814 \\ 
  9 & F      & Fluorine         & -99.8062$\pm$0.0001  & -99.7339  & -99.406    &  -99.409  &  -99.411 \\ 
 10 & Ne     & Neon             & -129.0507$\pm$0.0059 & -128.9376 & -128.542   & -128.545  & -128.547 \\ 
 11 & Na     & Sodium           & -162.4283$\pm$0.0032 & -162.2546 & -161.852   & -161.857  & -161.859 \\ 
 12 & Mg     & Magnesium        & -200.3100$\pm$0.0033 & -200.053  & -199.606   & -199.612  & -199.615 \\ 
 13 & Al     & Aluminum         & -242.7121$\pm$0.0037 & -242.346  & -241.868   & -241.873  & -241.877 \\ 
 14 & Si     & Silicon          & -289.8683$\pm$0.0037 & -289.359  & -288.845   & -288.851  & -288.854 \\ 
 15 & P      & Phosphorus       & -341.9464$\pm$0.0087 & -341.259  & -340.709   & -340.715  & -340.719 \\ 
 16 & S      & Sulfur           & -399.0351$\pm$0.0050 & -398.110  & -397.495   & -397.502  & -397.506 \\ 
 17 & Cl     & Chlorine         & -461.3813$\pm$0.0051 & -460.148  & -459.470   & -459.478  & -459.483 \\ 
 18 & Ar     & Argon            & -529.1122$\pm$0.0093 & -527.540  & -526.804   & -526.812  & -526.817 \\ 
 19 & K      & Potassium        & -601.9677$\pm$0.0515 & \hfil     & -599.150   & -599.159  & -599.165 \\ 
 20 & Ca     & Calcium          & -680.1022$\pm$0.0679 & \hfil     & -676.743   & -676.752  & -676.758 \\ 
 21 & Sc     & Scandium         & \hfil                & \hfil     & -759.718   & -759.728  & -759.736 \\ 
 22 & Ti     & Titanium         & \hfil                & \hfil     & -848.375   & -848.397  & -848.407 \\ 
 23 & V      & Vanadium         & \hfil                & \hfil     & -942.852   & -942.876  & -942.886 \\ 
 24 & Cr     & Chromium         & \hfil                & \hfil     & -1043.334  & -1043.350 & -1043.360 \\ 
 25 & Mn     & Manganese        & \hfil                & \hfil     & -1149.848  & -1149.860 & -1149.870 \\ 
 26 & Fe     & Iron             & \hfil                & \hfil     & -1262.425  & -1262.440 & -1262.450 \\ 
 27 & Co     & Cobalt           & \hfil                & \hfil     & -1381.376  & -1381.410 & -1381.420 \\ 
 28 & Ni     & Nickel           & \hfil                & \hfil     & -1506.828  & -1506.860 & -1506.870 \\ 
 29 & Cu     & Copper           & \hfil                & \hfil     & -1638.938  & -1638.950 & -1638.960 \\ 
 30 & Zn     & Zinc             & \hfil                & \hfil     & -1777.820  & -1777.830 & -1777.850 \\ 
 31 & Ga     & Gallium          & \hfil                & \hfil     & -1923.235  & -1923.250 & -1923.260 \\ 
 32 & Ge     & Germanium        & \hfil                & \hfil     & -2075.335  & -2075.350 & -2075.360 \\ 
 33 & As     & Arsenic          & \hfil                & \hfil     & -2234.215  & -2234.230 & -2234.240 \\ 
 34 & Se     & Selenium         & \hfil                & \hfil     & -2399.844  & -2399.860 & -2399.870 \\ 
 35 & Br     & Bromine          & \hfil                & \hfil     & -2572.416  & -2572.430 & -2572.440 \\ 
 36 & Kr     & Krypton          & \hfil                & \hfil     & -2752.029  & -2752.040 & -2752.050 \\ 
\hline
\end{tabular}
\end{center}
\caption{\label{table::atomic_energies} Total energies in Hartree for the atom series from Helium to Krypton. 
Two columns with experimental data are given, EXP$^{(a)}$  and EXP$^{(b)}$, for details see \appendixname~\ref{appendix:exp_data}.
 Other columns show the total energies obtained by various methods (SIF, OEP and \HF) within the exchange-only mode. 
The data are plotted in Fig.~\ref{pic:atomic_energies}. 
 In general, the SIF energies for the atom series lie slightly above the OEP ones, and both of them are higher than the \HF results. 
The differences are discussed in the text.}
\end{table*}
\endgroup 

Experimental data for the total energy can be obtained by summing all ionization potentials of the corresponding atom, a process whose difficulty increases prohibitively with atomic number, $Z$. 
Published results for the total energy are often characterized by a number of corrections to account for zero-point motion and relativistic effects. 
The resulting rather unclear situation for the total energies is discussed in \appendixname~\ref{appendix:exp_data}. 
In any case, the experimental values of the ground-state energies lie the lowest compared with those of HF, OEP and SIF in correspondingly increasing order.
We discuss this order in more detail in the following section.

\section{Differences between HF, OEP and the SIF Methods}
\label{sec::oep_sif}

\subsection{Comparison with Hartree-Fock}
It has become customary in the literature to compare the results of an exchange-only implementation of DFT to those of the \HF method, and we follow this practice. 

The \HF method relies on the direct calculation of a determinantal wave function to minimize the energy of an interacting system. 
The single-particle orbitals defining the determinant are the basic variables of the method, and their unrestricted variation provides a path to the lowest energy obtained in this procedure. 
The method, however, does not provide a path to the determination of the ground state of an interacting system.

By contrast, the basic variable in DFT is the density. %
In order to minimize the energy the functional derivative with respect to the density is  taken, and evaluated at that particular density formed by the \KS orbitals.  
In an exchange-only implementation of the theory, the restriction of the orbitals to reproduce the total energy minimizing energy density is expected to lead to higher values of the total energy compared with those of the \HF method and of course the ground state energy. 
This, indeed, is the case in all calculations reported here, see Table~\ref{table::atomic_energies}. 
The restrictions can also be discussed in terms of the potential. 
Within the \KS scheme, the potential, see Eq.~(\ref{Vs1}),  is unique to all orbitals, where in \HF it is orbital-dependent. This allows \HF to explore a broader space and may lead to lower energies.

\subsection{Comparison with OEP}

Since the most widely used method for obtaining a potential from an energy functional written explicitly in terms of the orbitals is the OEP we compare our results with the ones obtained in that method.
As is evident from Table~\ref{table::atomic_energies} and Figs.~\ref{pic:atomic_energies} and \ref{pic:errors}, the SIF total energies are systematically higher that those of the OEP.  Indeed, we find the consistent relationship  $E_{\rm HF} \le E_{\rm OEP} \le E_{\rm SIF} <E_{\rm LDA}$. As is particularly evident from Fig.~\ref{pic:errors}, the energy differences between both OEP and SIF relative to HF are small, as then are the differences between OEP and SIF, particularly when judged on the scale of the energy differences between either of these and  the corresponding LDA energies. In absolute terms, the root-mean-square deviations of the OEP, SIF and LDA energies from the HF energies, taken over the 36-atom set of Table~\ref{table::atomic_energies}, are $~8.22\times10^{-3}, ~2.04\times10^{-2}, ~1.03\times10^{0}$~H respectively.  Again, the slightly lower energies found in the OEP are indicative of some additional variational freedom contained in that approach relative to the current implementation of SIF. This despite the fact that both aim to evaluate the same functional derivative. How this difference is related to the EEB ansatz remains a matter for ongoing research.

\subsection{Alternative Method}
\label{sec:Sahni}
\begin{figure}[htp]
\usepictures{
\begin{center}
    \includegraphics[width=0.45\textwidth]{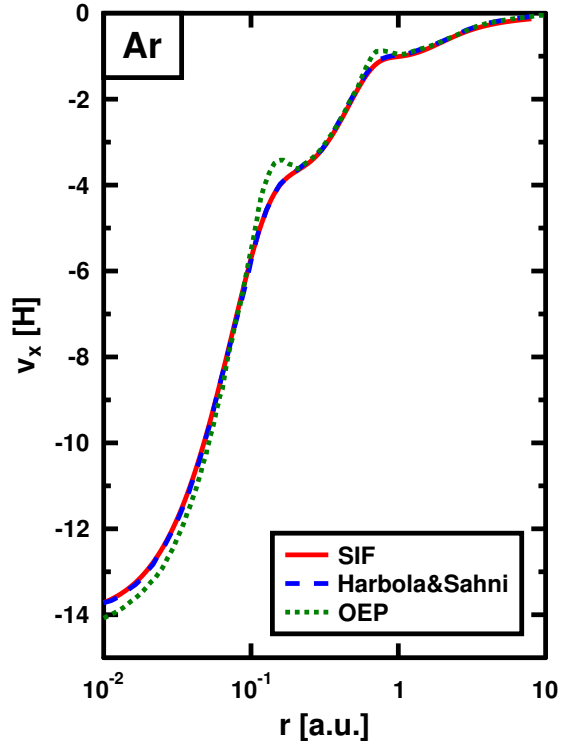}
 \end{center}
}
 \caption{Comparison between the exchange potential of the Argon atom derived by the methods used in this paper (SIF, red, solid), 
   the approach of Harbola and Sahni~\cite{HarbolaSahni1989_Ar} (dashed, blue), %
   and full OEP (dotted, green).}
 \label{pic:plot_Ar_sahni_sif_oep_yang}
\end{figure}

For special cases, e.g., closed shell atomic systems and generally spherical charge densities, Harbola and Sahni~\cite{HarbolaSahni1989_Ar} have shown 
that the exchange potential can be obtained through the use of classical electrostatics.  
Notably, their  method does not require the calculation of the susceptibility or its inverse and is independent of both SIF and OEP. 
For the case of Ar, the exchange potential found using their procedure precisely matches that which we 
find using the SIF approach (see Figure~\ref{pic:plot_Ar_sahni_sif_oep_yang}). As is also evident, both curves deviate somewhat from OEP, in
particular neither the Harbola and Sahni results nor SIF show the, so called, intershell structure of OEP.

In %
further work~\cite{Sahni_atoms_He_Rn}, Harbola and Sahni determined the ground-state energies of atomic systems from He to Rn. 
For closed shell systems, where densities are spherical and their method is applicable, their results coincide with ours as is clearly seen in
Table~\ref{table::atomic_energies} and Fig.~\ref{pic:plot_Ar_sahni_sif_oep_yang}. 
Of course, the SIF approach, unlike that Harbola and Sahni, can be applied quite generally to the non-spherical case.

\section{Conclusions}
\label{sec:conclusions}

The developments in the paper sprang from a particularly simple observation: The functional derivative of the Hartree term with respect to the density is facilitated by the explicit appearance of the density in the integrand. 
The same ease would materialize were the orbitals in the exchange term given explicitly in terms of the density in a differentiable functional form. 
The expansion of each orbital in the equidensity basis is particularly convenient for satisfying this requirement.

Based on this expansion, we have proposed a novel treatment for the Coulomb energy to be used in implementations of 
the \KS formulation of DFT within a local approximation.
In this treatment, the Coulomb energy is expressed in terms of the pair density constructed 
from the single-particle orbitals arising from the solution of the \KS equations, 
thus avoiding by construction self-interaction effects. 
The pair density, in turn, is written in closed-form explicitly in terms of the density. 
This is accomplished through an expansion of the single-particle orbitals in terms of the equidensity 
basis whose elements are written explicitly in terms of the density. The resulting equations for the functional derivative of the 
pair density can then also be written close form. Furthermore, they can evaluated straightforwardly provided that we neglect any possible 
implicit density dependence 
of the expansion coefficients of the equidensity basis in  Eq.~(\ref{ZM:FDerivativeOrbital})
We have then demonstrated how infinite sums appearing in the formalism can be replaced by quantities containing 
spatial derivatives and can thus evaluate the sums (Eqs.~(\ref{ZM:1D:spatialTrick}) and (\ref{ZM:def:QQQ})) to infinite order leading to a
computationally simple approach.

We have shown analytically that the approach recovers the exact expression for the exchange potential of two electron systems. Numerical studies 
of tunable interacting one-dimensional square well potential model are used to illustrate details of the inner workings of the approach and to show how it performs as a function of the interaction strength.

For real systems, we have shown results of the total energy of the atoms He through Kr and compared them with those obtained using HF, OEP, the approach of Harbola an Sahni, and standard LDA (including correlation). We find rather close agreement with the results of OEP for the total energies, charge densities and corresponding exchange potentials. Interestingly, for systems where that method is applicable we find very precise agreement with Harbola and Sahni, with neither of these exhibiting the so called inter-shell structure seen in the exchange potential of OEP. In terms of the energy the we find the hierarchy $E_{\rm HF} \le E_{\rm OEP} \le E_{\rm SIF}$. The fact that OEP energies are systematically slightly lower than those of SIF we attribute to the presence of additional variational freedom in that method compared with the current implementation of SIF.

The formalism presented in this paper can immediately be applied to all non-periodic systems, e.g. molecules. It is also likely that it can also be of considerable value for  solids. 
Increasingly, DFT calculations based on exchange correlation functionals (e.g., hybrid functionals) that contain some element of exact exchange are being used to study wide classes of systems (e.g., metallic oxides) where earlier generation functionals, e.g., LDA, GGA, fail to give a good overall description of the electronic structure and bonding. Currently such "hybrid-functional" calculations are very computationally demanding which 
severely limits the systems sizes and complexity, and time scales in $\it{ab initio}$ molecular dynamics, where they can be applied. Given that the approach presented in this paper yields expressions that are of closed form, they have the potential to be easily implemented and to significantly reduce the  computational demands relative to existing  OEP implementations. This in turn could allow hybrid-functional type methods to be applied to systems sizes currently only accessible using conventional LDA, GGA etc.
Finally, we point out that the presented procedure can generally applied to any orbital dependent functional in the same way as demonstrated here on the exchange energy.  

\section{Acknowledgments}
We are grateful to Stefan Kurth and Weitao Yang for providing their results of OEP calculations with which ours could be compared. 
We also gratefully acknowledge comments by X.-G. Zhang, Klaus Capelle, Hardy Gross and Viraht Sahni. 
We thank as well Patrice~E.~A. Turchi and Chris Billman for a careful and critical reading of the manuscript. 
The work at LLNL is supported by the U.S. DOE under Contract DE-AC52-07NA27344 with LLNS, LLC (MD, AG).
Research at ORNL is sponsored by the
Division of Materials Sciences and Engineering, Office of Basic
Energy Sciences (MD, GMS), and the 
{\it Center for Defect Physics in Structural Materials (CDP)}, 
an Energy Frontier Research Center funded by the U.S. Department of Energy, 
Office of Science, Office of Basic Energy Sciences (DMN, GMS).

\appendix
\section{Calculation of the integrals $I^{ij}$} \label{appendix:Iij}
In spherical coodinates, the integrals $I^{ij}$ defined in Eq.~(\ref{ZM:eq:def:Iij}) can be evaluated using the well known formula based on an expansion of~$U$ into spherical harmonics,
\begin{equation}
 \frac{1}{\left|\vec{r}-\vrs \right|}=\sum_{L=(l,m)} \frac{4\pi}{2l+1}\, \frac{r^l_<}{r^{l+1}_>} \,
  Y_L(\hat{r}) \, Y_L^{*}(\hat{r}^\prime), %
\end{equation}
and
\begin{eqnarray}
 &&\int_0^\infty {\rm d}r^{\prime}\, r^{\prime\, 2} \, g(r^{\prime}) \frac{r^l_<}{r^{l+1}_>} \hspace{5cm}  \nonumber \\
 & =&
  \frac{1}{r^{l+1}} \int_0^r {\rm d}r^{\prime}\, r^{\prime\, 2} \, r^{\prime \,l} g(r^{\prime}) \nonumber \\ &&+
  r^{l} \int _r^\infty {\rm d}r^{\prime}\,r^{\prime\, 2} \,  r^{\prime \,-l-1} \,g(r^{\prime}). %
\end{eqnarray}
In particular for a spherically symmetric function $\bar{g}$ we get,
\begin{equation}
\int{ \frac{\bar{g}(\vec{r}^{\prime})}{\left|\vec{r}-\vec{r}^{\prime} \right|}   {\rm d}\vrs} = \frac{4\pi}{r} \int_{0}^{r} \bar{g}(r^{\prime}) \, r^{\prime 2} \, {\rm d} r^{\prime} +
4\pi \int_{r}^{\infty} \bar{g}(r^{\prime}) \, r^{\prime} \, {\rm d} r^{\prime}.
\end{equation}
If the orbitals are of the form $\orb=R_L(r)\,Y_L(\theta,\phi)$ the angular part can be integrated 
analytically using the Gaunt coefficients,
\begin{equation}
 C_{L^{\prime}L^{\prime\prime}}^L=\int {\rm d}\hat{r}\;
     Y_{L^{\prime}}(\hat{r}) \;
     Y_{L^{\prime\prime}}^{*}(\hat{r}) \;
     Y_{L}(\hat{r})
\qquad L=(l,m).
 \label{eq:gaunts:def}
\end{equation}
Then the integrals $I^{ij}$  have the form,
\begin{eqnarray}
 I^{ij}(\vec{r}_1)&=&\sum_{L}^{L_{max}} R^{ij}_L(r_1) \, 
  Y_L(\theta_1,\phi_1)\\
&& \quad \mbox{with}\; l_{max}=2 \max(l_i,l_j)\;\mbox{and} \nonumber\\
  R^{ij}_L(r_1) &=& R_i^{*}(r_1)\,R_j(r_1)\, \left( C^{L}_{L_i,L_j} \right)^{*},
\end{eqnarray}
where, depending on the symmetry, some coefficients $R^{ij}_{L}$ might vanish.
Since the functional derivatives of $\vec{R}$ contain $\delta$ distributions 
it will be difficult to expand them into a spherical harmonics basis. 
We find it more useful to evaluate the integrals $I^{ij}$ 
in Cartesian coordinates and  perform the remaining integrals to obtain the potential.%
\section{Deriatives of the transformed coordinates $\vec{R}$}
\label{FDEquiD}
\subsection{Spatial Deriatives}
\label{ZM:Appendix:fDerivativeSpacial}
For completeness, we provide all spatial derivatives of the functionals $\vec{R}(n(\vec{r},[n])$ defined in Eqs.~(\ref{ZM:R123}).
\begin{eqnarray}
\beta_{11}:= \frac{\partial R_{1}}{\partial x}&=& 2\pi\,\frac{n(\vec{r})}{N(y,z)} \label{ZM:beta11}\\
\beta_{12}:=\frac{\partial R_{1}}{\partial y}
&=& \frac{1}{N(y,z)} \Bigg[2\pi \int_{-\infty}^x{\rm d}x^\prime \frac{\partial}{\partial y} n(x^\prime,y,z) \nonumber \\
 && -R_{1}\int_{-\infty}^{\infty}{\rm d}x^\prime \frac{\partial}{\partial y} n(x^\prime,y,z)  \Bigg]\label{ZM:beta12}\\
\beta_{13}:=\frac{\partial R_{1}}{\partial z}
&=& \frac{1}{N(y,z)} \Bigg[2\pi \int_{-\infty}^x{\rm d}x^\prime \frac{\partial}{\partial z} n(x^\prime,y,z) \nonumber \\
 &&-R_{1}\,\int_{-\infty}^{\infty}{\rm d}x^\prime \frac{\partial}{\partial z} n(x^\prime,y,z)  \Bigg]\label{ZM:beta13}
\end{eqnarray}
\begin{eqnarray}
\beta_{21}:= \frac{\partial R_{2}}{\partial x}&=& 0 \label{ZM:beta21}\\
\beta_{22}:=\frac{\partial R_{2}}{\partial y}&=& 2\pi \frac{N(y,z)}{N(z)} \label{ZM:beta22} \\
\beta_{23}:= \frac{\partial R_{2}}{\partial z}
&=&  \frac{1}{N(z)} \Bigg[ 2\pi \int_{-\infty}^y{\rm d}y^\prime\int_{-\infty}^{\infty}{\rm d}x^\prime \frac{\partial}{\partial z}n(x^\prime,y^\prime,z)\nonumber \\
    &&- R_2 \, \int_{-\infty}^{\infty}{\rm d}y^\prime\int_{-\infty}^{\infty}{\rm d}x^\prime \frac{\partial}{\partial z}n(x^\prime,y^\prime,z) \Bigg]\label{ZM:beta23}
\end{eqnarray}
\begin{eqnarray}
\beta_{31}:= \frac{\partial R_{3}}{\partial x}&=& 0 \label{ZM:beta31}\\
\beta_{32}:= \frac{\partial R_{3}}{\partial y}&=& 0 \label{ZM:beta32}\\
\beta_{33}:= \frac{\partial R_{3}}{\partial z}&=&  2\pi \frac{N(z)}{N} \label{ZM:beta33}
\end{eqnarray}
We see  that the Jacobian is
\begin{equation}
  \det \beta = \det \left| \begin{array}{ccc}
  \frac{\partial R_1}{\partial x} & \frac{\partial R_1}{\partial y} & \frac{\partial R_1}{\partial z} \\
                                0 & \frac{\partial R_2}{\partial y} & \frac{\partial R_2}{\partial z} \\
                                0 &                                 0 & \frac{\partial R_3}{\partial z} 
 \end{array} \right|= \frac{(2\pi)^3}{N} n(\vec{r}),
\end{equation}
and in general $\det \beta = \frac{(2\pi)^d}{N} n(\vec{r})$ where $d$ is the dimension of the system.

\subsection{Functional Derivatives of the $\vec{R}$}\label{ZM:Appendix:fDerivativeR}
Straightforward functional differentiations lead to the expressions: 
\begin{widetext}

\begin{eqnarray}
 \frac{\delta R_1(x,y,z,[n^\sigma])}{\delta n(\vec{r}^{\prime\prime})} 
&=& 2\pi  \frac{\delta}{\delta n(\vec{r}^{\prime\prime})}\,
   \frac{\int_{-\infty}^{x}  {\rm d}x^{\prime} \, n(x^\prime,y,z,[n^\sigma])}{ \int_{-\infty}^{\infty}  {\rm d}x^{\prime} \, n(x^\prime,y,z,[n^\sigma]) } \nonumber \\
&=& 2\pi \frac{
   \int_{-\infty}^{x} {\rm d}x^\prime 
   \delta(x^\prime-x^{\prime\prime})
   \delta(y-y^{\prime\prime})
   \delta(z-z^{\prime\prime})
   } 
   {  \int_{-\infty}^{\infty}  {\rm d}x^{\prime} \, n(x^\prime,y,z,[n^\sigma])   } \nonumber \\
&&
- 2\pi 
    \int_{-\infty}^{\infty} {\rm d}x^\prime 
   \delta(x^\prime-x^{\prime\prime})
   \delta(y-y^{\prime\prime})
   \delta(z-z^{\prime\prime}) %
 \nonumber \\ && \times
   \frac{\int_{-\infty}^{x}  {\rm d}x^{\prime} \, n(x^\prime,y,z,[n^\sigma])}{\left[  \int_{-\infty}^{\infty}  {\rm d}x^{\prime} \, n(x^\prime,y,z,[n^\sigma])   \right]^2}   \nonumber \\
&=& \frac{\delta(y-y^{\prime\prime}) \delta(z-z^{\prime\prime}) } { N(y,z,[n^\sigma]) } %
\bigg[ 2\pi\, \Theta(x-x^{\prime\prime}) -R_1(x,y,z,n[\sigma])  \bigg], \nonumber
\end{eqnarray}
\begin{eqnarray}
\frac{\delta R_2(y,z,[n^\sigma])}{\delta n(\vec{r}^{\prime\prime})} %
&=& 2\pi  \frac{\delta}{\delta n(\vec{r}^{\prime\prime})}\,
   \frac{   \int_{-\infty}^{y}       {\rm d}y^{\prime} \int_{-\infty}^{\infty}  {\rm d}x^{\prime} \, n(x^\prime,y^\prime,z,[n^\sigma])}
         {   \int_{-\infty}^{\infty}  {\rm d}y^{\prime} \int_{-\infty}^{\infty}  {\rm d}x^{\prime} \, n(x^\prime,y^\prime,z,[n^\sigma]) }\nonumber \\
&=& 2\pi \frac{
   \int_{-\infty}^{y}       {\rm d}y^{\prime} \int_{-\infty}^{\infty}  {\rm d}x^{\prime} 
   \delta(x^\prime-x^{\prime\prime})
   \delta(y^\prime-y^{\prime\prime})
   \delta(z-z^{\prime\prime})
   } 
   {   \int_{-\infty}^{\infty}  {\rm d}y^{\prime} \int_{-\infty}^{\infty}  {\rm d}x^{\prime} \, n(x^\prime,y^\prime,z,[n^\sigma])  } \nonumber \\
&&- 2\pi 
   \int_{-\infty}^{\infty}       {\rm d}y^{\prime} \int_{-\infty}^{\infty}  {\rm d}x^{\prime} 
   \delta(x^\prime-x^{\prime\prime})
   \delta(y^\prime-y^{\prime\prime})
   \delta(z-z^{\prime\prime})
 \nonumber \\ && \times   
\frac{   \int_{-\infty}^{y}       {\rm d}y^{\prime} \int_{-\infty}^{\infty}  {\rm d}x^{\prime} \, n(x^\prime,y^\prime,z,[n^\sigma])       }
  {\left[  \int_{-\infty}^{\infty}  {\rm d}y^{\prime} \int_{-\infty}^{\infty}  {\rm d}x^{\prime} \, n(x^\prime,y^\prime,z,[n^\sigma])  \right]^2}   \nonumber \\
&=& \frac{\delta(z-z^{\prime\prime}) } { N(z,[n^\sigma]) }  \bigg[ 2\pi\, \Theta(y-y^{\prime\prime}) -R_2(y,z,n[\sigma])  \bigg], \nonumber
\end{eqnarray}
and
\begin{eqnarray}
\frac{\delta R_3(z,[n^\sigma])}{\delta n(\vec{r}^{\prime\prime})} %
&=& 2\pi  \frac{\delta}{\delta n(\vec{r}^{\prime\prime})}\,
   \frac{   \int_{-\infty}^{z}       {\rm d}z^{\prime} \int_{-\infty}^{\infty}  {\rm d}y^{\prime} \int_{-\infty}^{\infty}  {\rm d}x^{\prime}  \, n(x^\prime,y^\prime,z^\prime,[n^\sigma])}
        {   \int_{-\infty}^{\infty}  {\rm d}z^{\prime} \int_{-\infty}^{\infty}  {\rm d}y^{\prime} \int_{-\infty}^{\infty}  {\rm d}x^{\prime}  \, n(x^\prime,y^\prime,z^\prime,[n^\sigma]) }\nonumber \\
&=& 2\pi \frac{
   \int_{-\infty}^{z}  {\rm d}z^{\prime}       \int_{-\infty}^{\infty}  {\rm d}y^{\prime} \int_{-\infty}^{\infty}  {\rm d}x^{\prime} 
   \delta(x^\prime-x^{\prime\prime})
   \delta(y^\prime-y^{\prime\prime})
   \delta(z^\prime-z^{\prime\prime})
   } 
   {   \int_{-\infty}^{\infty}  {\rm d}z^{\prime}  \int_{-\infty}^{\infty}  {\rm d}y^{\prime} \int_{-\infty}^{\infty}  {\rm d}x^{\prime} \, n(x^\prime,y^\prime,z^\prime,[n^\sigma])  } \nonumber \\
&&- 2\pi 
   \int_{-\infty}^{\infty}       {\rm d}z^{\prime}    \int_{-\infty}^{\infty}  {\rm d}y^{\prime}  \int_{-\infty}^{\infty}  {\rm d}x^{\prime} 
   \delta(x^\prime-x^{\prime\prime})
   \delta(y^\prime-y^{\prime\prime})
   \delta(z^\prime-z^{\prime\prime})
 \nonumber \\ && \times 
  \frac{   \int_{-\infty}^{z}       {\rm d}z^{\prime}  \int_{-\infty}^{\infty}  {\rm d}y^{\prime} \int_{-\infty}^{\infty}  {\rm d}x^{\prime}  \, n(x^\prime,y^\prime,z^\prime,[n^\sigma])       }
  {\left[ \int_{-\infty}^{\infty}  {\rm d}z^{\prime}  \int_{-\infty}^{\infty}  {\rm d}y^{\prime} \int_{-\infty}^{\infty}  {\rm d}x^{\prime} \, n(x^\prime,y^\prime,z^\prime,[n^\sigma])  \right]^2}   \nonumber \\
&=& \frac{1} { N^\sigma }  \bigg[ 2\pi\, \Theta(z-z^{\prime\prime}) -R_3(z,n[\sigma])  \bigg].\nonumber
\end{eqnarray}

\end{widetext}%
\section{Spatial Derivatives and their Connection to Infinite Sums}
\label{appendix:spacial_trick}
The use of the equidensity basis requires the evaluation of the rather difficult infinite sum in 
Eq.~(\ref{ZM:FDerivativeOrbital}) whose value depends on the number of terms taken into account. 
The following considerations allow the sum to be carried out to infinite order.

First, we discuss the one-dimensional case. We are interested in evaluating the sum,
\begin{equation}
 \sum_{k} a_{k} \, i\,k  \,  \phi_{k}(x,[n^\sigma]) = \sum_{k} a_{k} \, i\,k \sqrt{\frac{n^\sigma(x)}{N^\sigma}} e^{i\,k\,q(x,[n^\sigma])},
\end{equation}
where the orbitals $\orb(x)$ are expressed in the form (where for simplicity of notation we omit the spin superscript on the orbitals),
\begin{equation}
\orb(x)=
\sum_{k} a_{k} \, \phi_{\vec{k}}=
\sum_{k} a_{k} \sqrt{\frac{n^\sigma(x)}{N^\sigma}} e^{i\,k\,q(x,[n^\sigma])}.
\end{equation}
From the definition, 
\beq
q(x,[n^\sigma])=\frac{2\pi}{N^\sigma}\int_{x_0}^{x} n^\sigma(x^{\prime})\,{\rm d}x^{\prime}
\label{qx.a1}
\eeq
we have,
\beq
 \frac{\delta q(x,[n^\sigma])}{\delta n^\sigma(x^{\prime})} = \frac{2\pi}{N^\sigma}\,\Theta(x-x^{\prime}),
 \label{DerQx.1}
 \eeq
 and
 \beq
\frac{{\rm d}q(x,[n^\sigma])}{{\rm d}x}=\frac{2\pi}{N^\sigma} n^\sigma(x).
\eeq
The spatial derivative is given by
\begin{eqnarray}
 f^{\prime}&=&\frac{{\rm d}f(x)}{{\rm d}x} \nonumber \\
&=& \sum_{k} a_{k} \frac{{\rm d}}{{\rm d}x} \left( \sqrt{\frac{n^\sigma(x)}{N^\sigma}} e^{i\,k\,q(x,[n^\sigma])}   \right) \\ \nonumber
&=& \sum_{k}  \Bigg[ 
     \frac{1}{\sqrt{N^\sigma}} a_k\, \frac{1}{2 \sqrt{n^\sigma(x)}} \frac{dn^\sigma(x)}{dx} e^{i\,k\,q(x,[n^\sigma])}\\ \nonumber
   &+& i\,k\,a_k \sqrt{\frac{n^\sigma(x)}{N^\sigma}} e^{i\,k\,q(x,[n^\sigma])} \frac{dq(x)}{dx} \Bigg] \\ \nonumber 
&=& \frac{n^{\sigma\prime}(x)}{2 n^\sigma(x)} \underbrace{\sum_{k} a_k \sqrt{\frac{n^\sigma(x)}{N^\sigma}} e^{i\,k\,q(x,[n^\sigma])}}_{f(x)}\\
  &+& \underbrace{\frac{2\pi}{N^\sigma} n^\sigma(x)}_{q^{\prime}} \sum_{k} a_k \,i\,k \sqrt{\frac{n^\sigma(x)}{N^\sigma}} e^{i\,k\,q(x,[n^\sigma])},
\end{eqnarray}
from which it follows that
\begin{eqnarray}
\nonumber &\left.\right.& \sum_{k} a_k \,i\,k \sqrt{\frac{n^\sigma(x)}{N^\sigma}} e^{i\,k\,q(x,[n^\sigma])} \\ \nonumber
&=& \frac{N^\sigma}{2\pi\,n^\sigma(x)} \, \frac{{\rm d}f(x)}{{\rm d}x} - \frac{N^\sigma}{4\pi\,[n^\sigma(x)]^2}\, f(x) \,\frac{{\rm d}n^\sigma(x)}{{\rm d}x} \\ %
&=& \frac{N^\sigma}{2\pi\,n^\sigma(x)} \left[f^{\prime} -\frac{1}{2\,n^\sigma(x)}  \,f(x)\, n^{\sigma\prime}\right]\\
&=& \frac{N^\sigma}{2\pi\,n^\sigma(x)} \left[ \sqrt{n^\sigma(x)} \left( \frac{f(x)}{\sqrt{n^\sigma(x)}}\right)^\prime\right] \\
&=& \frac{N^\sigma}{2\pi} \frac{1}{\sqrt{n^\sigma(x)}} \left( \frac{f(x)}{\sqrt{n^\sigma(x)}}\right)^\prime
\label{1DQ.1}
\end{eqnarray}
The spatial derivatives with respect to $x$ (marked by primes) of the orbitals and the density can be calculated numerically.

It now follows that it is no longer necessary to construct explicitly the equidensity basis, or the expansion coefficients, $a_{\vec{k}}$.
The three-dimensional generalization of the one-dimensional results is straightforward.
The components of the gradient of an orbital, $\orb$, decomposed into the equidensity basis are given by
\begin{eqnarray}
 \nabla_{\alpha}\orb &=& \sum_{\vk} a^\sigma_{\vk} \frac{\partial}{\partial\alpha} \left( \sqrt{\frac{n^\sigma(\vec{r})}{N^\sigma}}  e^{i\,\,\vk\cdot \vec{R}(\vec{r},[n^\sigma])}  \right)\\
&=& \frac{1}{2 n^\sigma(\vec{r})}  \frac{\partial n^\sigma(\vec{r})}{\partial\alpha} \orb(\vec{r}) \nonumber \\
&&+ i \sum_{\vk} a^\sigma_{\vk}
 \phi^\sigma_{\vk}(\vec{r},[n^\sigma]) \,
\left(\frac{\partial \vec{R}(\vec{r},[n^\sigma])}{\partial\alpha}\cdot \vk  \right),
\label{ZM:gradientOrb}
\end{eqnarray}
with $\alpha=\{x,y,z\}$. \\
From Eqs. (\ref{ZM:gradientOrb}), (\ref{ZM:beta11}), (\ref{ZM:beta21}) and (\ref{ZM:beta31}) we obtain
\begin{eqnarray}
  \nabla_x\orb 
  &=& \frac{1}{2n^\sigma(\vec{r})} \frac{\partial n^\sigma(\vec{r})}{\partial x} \orb(\vec{r}) \nonumber\\
  & &+ \sum_{\vk} a^\sigma_{\vk} \phi^\sigma_{\bf k}(\vec{r}) (i k_1) 
\underbrace{\left( 2\pi\frac{n^\sigma(\vec{r})}{N^\sigma(y,z)} \right)}_{\beta_{11}}, 
\end{eqnarray}
and
\begin{eqnarray}
Q_x^\sigma=\sum_{\vk} a^\sigma_{\vk} \phi^\sigma_{\bf k}(\vec{r}) (i k_1) 
&=&\frac{N^\sigma(y,z)}{2\pi n^\sigma(\vec{r})} \nonumber\\
   &&\hspace{-3cm}\times\bigg[ 
   \nabla_x\orb-\frac{\orb(\vec{r})}{2n^\sigma(\vec{r})} \frac{\partial n^\sigma(\vec{r})}{\partial x}
   \bigg].%
\end{eqnarray}
In the same way, using Eqs.~(\ref{ZM:gradientOrb}), (\ref{ZM:beta12}), (\ref{ZM:beta22})  
and (\ref{ZM:beta32})
\begin{eqnarray}
 \nabla_y\orb 
&=& \frac{1}{2n^\sigma(\vec{r})} \frac{\partial n^\sigma(\vec{r})}{\partial y} \orb(\vec{r}) \nonumber\\
    &&+ \sum_{\vk} a^\sigma_{\vk} \phi^\sigma_{\bf k}(\vec{r}) (i k_1) \beta_{12}^\sigma\nonumber\\
    &&+ \sum_{\vk} a^\sigma_{\vk} \phi^\sigma_{\bf k}(\vec{r}) (i k_2)  
\underbrace{2\pi \frac{N^\sigma(y,z)}{N^\sigma(z)}}_{\beta_{22}},
\end{eqnarray}
we obtain the expression,
\begin{eqnarray}
Q_y^\sigma=\sum_{\vk} a^\sigma_{\vk} \phi^\sigma_{\bf k}(\vec{r}) (i k_2) 
&=&  \frac{N^\sigma(z)}{2\pi\,N^\sigma(y,z)}  \nonumber\\
    &&\hspace{-3cm}\times   \bigg[  \nabla_y\orb 
        - \frac{\orb(\vec{r})}{2 n^\sigma(\vec{r})} \frac{\partial n^\sigma(\vec{r})}{\partial y}
        - Q_x^\sigma \beta_{12}^\sigma
     \bigg].
\end{eqnarray}
Finally using Eqs. (\ref{ZM:gradientOrb}), (\ref{ZM:beta13}), (\ref{ZM:beta23}) and (\ref{ZM:beta33}), we find,
\begin{eqnarray}
 \nabla_z\orb 
&=& \frac{1}{2n^\sigma(\vec{r})} \frac{\partial n^\sigma(\vec{r})}{\partial z} \orb(\vec{r}) \nonumber\\
    && + \sum_{\vk} a^\sigma_{\vk} \phi^\sigma_{\bf k}(\vec{r}) (i k_1)      \beta_{13}^\sigma\nonumber\\
    && + \sum_{\vk} a^\sigma_{\vk} \phi^\sigma_{\bf k}(\vec{r}) (i k_2)      \beta_{23}^\sigma\nonumber\\
    && + \sum_{\vk} a^\sigma_{\vk} \phi^\sigma_{\bf k}(\vec{r}) (i k_3)\, 
\underbrace{2\pi \frac{N^\sigma(z)}{N^\sigma}}_{\beta_{33}},  
\end{eqnarray}
and
\begin{eqnarray}
 Q_z^\sigma&=&\sum_{\vk} a^\sigma_{\vk} \phi^\sigma_{\bf k}i(\vec{r}) (i k_3)
= \frac{N^\sigma}{2\pi\,N^\sigma(z)} \nonumber\\
 && \hspace{-1cm} \times   \bigg[
     \nabla_z\orb 
     -\frac{\orb(\vec{r})}{2n^\sigma(\vec{r})} \frac{\partial n^\sigma(\vec{r})}{\partial z}
     - Q_x^\sigma \beta_{13}^\sigma
     - Q_y^\sigma \beta_{23}^\sigma
    \bigg].
\end{eqnarray}

These results can be summarized in vector form,
\begin{eqnarray}
 \sum_{\vk} a_{\vk}^\sigma \phi^\sigma_{\bf k}(\vec{r}) (i \vk)
=:\left( \begin{array}{l} Q_x^\sigma \\Q_y^\sigma \\Q_z^\sigma \end{array} \right)= \hspace{3.5cm} \nonumber \\
 \left(  \begin{array}{ll}
\frac{N^\sigma(y,z)}{2\pi n^\sigma(\vec{r})} & \hspace{-8.1pt} \bigg[ 
   \nabla_x\orb-\frac{\orb(\vec{r})}{2n^\sigma(\vec{r})} \frac{\partial n^\sigma(\vec{r})}{\partial x}
   \bigg]  \\[.3cm]
\frac{N^\sigma(z)}{2\pi N^\sigma(y,z)}& \hspace{-8.1pt} \bigg[  \nabla_y\orb 
        - \frac{\orb(\vec{r})}{2n^\sigma(\vec{r})} \frac{\partial n^\sigma(\vec{r})}{\partial y}
        - Q_x^\sigma \beta_{12}^\sigma
     \bigg]  \\[.3cm]
  \frac{N^\sigma}{2\pi\,N^\sigma(z)}  & \hspace{-8.1pt} \bigg[
     \nabla_z\orb 
     -\frac{\orb(\vec{r})}{2n^\sigma(\vec{r})} \frac{\partial n^\sigma(\vec{r})}{\partial z}
     - Q_x^\sigma \beta_{13}^\sigma
     - Q_y^\sigma \beta_{23}^\sigma
    \bigg]
 \end{array} \right). \nonumber
\end{eqnarray}
\section{Numerical Evaluation of the Exchange Potential Components}\label{appendix:vx_details}

We split the exchange potential into different components with regard to Eqn.~(\ref{ZM:FDerivativeOrbitalQ}) and (\ref{eq:ZM:exchangePotential})
\begin{equation}
 v_x^{\sigma}(\vec{r}^{\prime}) =
 v_x^{\sigma [\mbox{1}]} (\vec{r}^{\prime})
+v_x^{\sigma [\mbox{Qx}]} (\vec{r}^{\prime})
+v_x^{\sigma [\mbox{Qy}]} (\vec{r}^{\prime})
+v_x^{\sigma [\mbox{Qz}]} (\vec{r}^{\prime}).
\end{equation}
In the following we show how to evaluate the different integrals to obtain the exchange potential, 
in cartesian coordinates,  which makes it easy to deal with integrals over Theta functions. 

The contribution from the first term of Eq.~(\ref{ZM:FDerivativeOrbitalQ}) takes the form,
\begin{eqnarray}
  v_x^{\sigma [\mbox{1}]} (\vec{r}^{\prime})
&=& -  2 \Re \sum_{ij} \delta_{\sigma_i,\sigma_j,\sigma} \nonumber \\
&& \hspace{0cm} \times \int {\rm d}\vec{r}_1 \, I^{ij}(\vec{r}_1) \,  
    \orb_j^{*}(\vec{r}_1) \, \orb_i(\vec{r}_1) \frac{\delta(\vec{r}_1-\vec{r}^{\prime})}{2\,n^\sigma(\vec{r}_1)} 
   \nonumber \\
&=& -   \frac{1}{n^\sigma(\vec{r}^{\prime})} \Re \sum_{ij} \delta_{\sigma_i,\sigma_j,\sigma} I^{ij}(\vec{r}^{\prime}) \, 
    \orb_j^{*}(\vec{r}^{\prime}) \, \orb_i(\vec{r}^{\prime}). \nonumber
\end{eqnarray}

The contribution that includes $\left( \frac{\partial R_1}{\partial\,n}\, Q_x\right)$ in the last term of Eq.~(\ref{ZM:FDerivativeOrbitalQ}), reads as follows
\begin{eqnarray}
&& v_x^{\sigma [\mbox{Qx}]} (\vec{r}^{\prime})= \nonumber \\
&& -  2 \Re \hspace{-.1cm}  \sum_{ij} \hspace{-.1cm}  \delta_{\sigma_i,\sigma_j,\sigma}\int {\rm d}\vec{r}\, I^{ij}(\vec{r}) \, \orb_j^{*}(\vec{r}) 
  \frac{\delta(y-y^{\prime}) \, \delta(z-z^{\prime})}{N^\sigma(y,z,[n])} \times \nonumber \\
   &&\Bigg[ 2\pi \, \Theta(x-x^{\prime}) - R_1(x,y,z,[n^\sigma]) \Bigg] \times 
     \frac{N^\sigma(y,z)}{2\pi n^\sigma(\vec{r})}  \times \nonumber \\ 
 && \bigg[ 
   \nabla_x\orb_i-\frac{\orb_i(\vec{r})}{2n^\sigma(\vec{r})} \frac{\partial n^\sigma(\vec{r})}{\partial x}
   \bigg] \nonumber \\
&=& - 2 \Re \hspace{-.1cm}  \int \hspace{-.1cm} {\rm d}x\,{\rm d}y\,{\rm d}z \,  \Bigg\{
     \Bigg[ 2\pi\, \Theta(x-x^{\prime})-R_1(x,y,z,[n^\sigma]) \Bigg] \hspace{-.1cm} \times \nonumber \\
   && \hspace{0.6cm} \Bigg[ \sum_{ij} \delta_{\sigma_i,\sigma_j,\sigma} \, I^{ij}(x,y,z) \, \orb_j^{*}(x,y,z) \, 
       \frac{Q_x^i(x,y,z)}{N^\sigma(y,z)} \Bigg] \times \nonumber \\
   && \hspace{0.9cm} \delta(y-y^{\prime}) \, \delta(z-z^{\prime}) \Bigg\}.
\end{eqnarray}
The integrals over $y$ and $z$ are trivial. Left over are two simple integral over $x$.

We can define:
\begin{eqnarray}
 A_{x}^{\sigma}(x,y,z)  &:=&\sum_{ij} \hspace{-.08cm} \delta_{\sigma_i,\sigma_j,\sigma} \, I^{ij}(x,y,z) \, \orb_j^{*}(x,y,z) 
       \frac{Q_x^i(x,y,z)}{N^\sigma(y,z)} \nonumber \\
 A_{y}^{\sigma}(x,y,z)  &:=&\sum_{ij} \hspace{-.08cm} \delta_{\sigma_i,\sigma_j,\sigma} \, I^{ij}(x,y,z) \, \orb_j^{*}(x,y,z) 
       \frac{Q_y^i(x,y,z)}{N^\sigma(z)}\nonumber \\
 A_{z}^{\sigma}(x,y,z)  &:=&\sum_{ij} \hspace{-.08cm} \delta_{\sigma_i,\sigma_j\sigma} \, I^{ij}(x,y,z) \, \orb_j^{*}(x,y,z) 
       \frac{Q_z^i(x,y,z)}{N^\sigma} , \nonumber
\end{eqnarray}
that leads to
\begin{eqnarray}
 && v_x^{\sigma [\mbox{Qx}]} (\vec{r}^{\prime}) = v_x^{\sigma [\mbox{Qx}]} (x^{\prime},y^{\prime},z^{\prime}) 
     \nonumber \\
 &=&   -2 \Re \Bigg[    \int_{x^{\prime}}^{\infty} {\rm d}x \,   2\pi\, A_{x}^{\sigma}(x,y^{\prime},z^{\prime}) \nonumber \\
    && \hspace{0cm} -\int_{-\infty}^{\infty}         {\rm d}x \,   R_1(x,y^{\prime},z^{\prime},[n^\sigma]) \, A_{x}^{\sigma}(x,y^{\prime},z^{\prime}) \Bigg].
\end{eqnarray}

The contribution from the term that includes $\left( \frac{\partial R_2}{\partial\,n}\, Q_y\right)$ in the last term of equation~(\ref{ZM:FDerivativeOrbitalQ}), reads as follows
\begin{eqnarray}
 && v_x^{\sigma [\mbox{Qy}]} (\vec{r}^{\prime}) = v_x^{\sigma [\mbox{Qy}]} (y^{\prime},z^{\prime})  \nonumber \\
 &=& - 2 \Re \int {\rm d}x\,{\rm d}y\,{\rm d}z \,  \Bigg\{
     \Bigg[ 2\pi\, \Theta(y-y^{\prime})-R_2(y,z,[n^\sigma]) \Bigg] \times \nonumber\\
   && \hspace{0.6cm} \Bigg[ \sum_{ij} \delta_{\sigma_i,\sigma_j,\sigma} \, I^{ij}(x,y,z) \, \orb_j(x,y,z) \, 
       \frac{Q_y^i(x,y,z)}{N^\sigma(z)} \Bigg] \times \nonumber \\
   && \hspace{0.9cm}  \delta(z-z^{\prime}) \Bigg\}    \\
  &=& -2 \Re \Bigg[ \int_{y^{\prime}}^{\infty} {\rm d}y \int_{-\infty}^{\infty} {\rm d}x \,
                2\pi\,A_{y}^{\sigma}(x,y,z^{\prime}) \nonumber \\
   &&\hspace{0.4cm}-\int_{-\infty}^{\infty}         {\rm d}y \int_{-\infty}^{\infty}         {\rm d}x \,
          R_2(y,z^{\prime},[n^\sigma]) \,A_{y}^{\sigma}(x,y,z^{\prime})\Bigg].
\end{eqnarray}

The contribution from the term that includes $\left( \frac{\partial R_3}{\partial\,n}\, Q_z\right)$ in the last term of equation~(\ref{ZM:FDerivativeOrbitalQ}), reads as follows
\begin{eqnarray}
&& v_x^{\sigma [\mbox{Qz}]} (\vec{r}^{\prime})= v_x^{\sigma [\mbox{Qz}]} (z^{\prime})  \nonumber \\
 &=& - 2 \Re \int {\rm d}x\,{\rm d}y\,{\rm d}z \,  \Bigg\{
     \Bigg[ 2\pi\, \Theta(z-z^{\prime})  -R_3(z,[n^\sigma]) \Bigg] \times \nonumber\\
   && \hspace{0.0cm} \Bigg[ \sum_{ij} \delta_{\sigma_i,\sigma_j\sigma} \, I^{ij}(x,y,z) \, \orb_j(x,y,z) \, 
       \frac{Q_z^i(x,y,z)}{N^\sigma} \Bigg]  \Bigg\} \nonumber \\
  &=& -2 \Re \Bigg[ \int_{z^{\prime}}^{\infty} {\rm d}z \int_{-\infty}^{\infty} {\rm d}y \int_{-\infty}^{\infty} {\rm d}x \,
         2\pi\,A_{z}^{\sigma}(x,y,z) \nonumber \\
    &&  \hspace{0.0cm} \underbrace{ -\int_{-\infty}^{\infty} {\rm d}z \int_{-\infty}^{\infty} {\rm d}y \int_{-\infty}^{\infty} {\rm d}x
           R_3(z,[n^\sigma])\,A_{z}^{\sigma}(x,y,z)  }_{\mbox{this is a constant, spin dependent}}  \Bigg] . \nonumber\\
\end{eqnarray}
\section{Functional Derivative of the Hartree term} \label{appendix:VHartree}
For the Hartree term it is easy to show that
\begin{equation}
\frac{\delta E_{\rm H}}{\delta n^{\up}(\vec{r})}=
\frac{\delta E_{\rm H}}{\delta n^{\down}(\vec{r})}=
\frac{\delta E_{\rm H}}{\delta n(\vec{r})}=v_{\rm H}(\vec{r}).
\end{equation}
Nevertheless, it still needs to be proven that the proposed procedure calculating the 
functional derivative with respect to the density leads to the exact same result, 
namely the Hartree potential.

For the spin density we have
\begin{eqnarray}
 n^{\sigma}(\vec{r})&=&\sum_i^{N^{\sigma}} \orb_i^{\sigma *}(\vec{r})\,\orb_i^\sigma (\vec{r}) \\
 \nabla_{\alpha}n^{\sigma}(\vec{r})&=&\sum^{N^\sigma}_i \Big[  \orb_i^{\sigma *}(\vec{r}) \; \nabla_{\alpha}\orb_i^\sigma(\vec{r}) + \orb_i^\sigma(\vec{r}) \; \nabla_{\alpha}\orb_i^{\sigma*}(\vec{r}) \Big] \nonumber \\
\end{eqnarray}
where $\alpha\in\{x,y,z\}$ and the sum runs over all orbitals of spin $\sigma$. 
In the spin polarized case there is no functional dependency of an orbital with spin $\sigma$ and the 
spin density with the opposite spin. 
\begin{equation}
\frac{\delta \orb_i^{\up}}{\delta n^{\down}}=0=\frac{\delta \orb_i^{\down}}{\delta n^{\up}}
\end{equation}
Now we can show that the  following expression vanishes (here shown for $\sigma=\up$):
\begin{eqnarray}
&&\hspace{-1cm} \sum_{i}^{\up only} \Bigg\{ \orb_i^{*} \Big[ 
   \nabla_{\alpha}\orb_i-\frac{\orb_i}{2n} \frac{\partial n^{\sigma}}{\partial \alpha} \Big] 
  + \orb_i \Big[ 
     \nabla_{\alpha}\orb_i^{*}-\frac{\orb_i^{*}}{2n} \frac{\partial n^{\sigma}}{\partial \alpha}
  \Big] \Bigg\} \label{ZM:eq:3D_simple1} \\
&=& \sum_i^{\up only} \Big[  \orb_i^{*} \; \nabla_{\alpha}\orb_i + 
 \orb_i(\vec{r}) \; \nabla_{\alpha}\orb_i^{*} \Big]  %
 -\nabla_{\alpha}n^{\sigma}\sum_i^{\up only} \frac{\orb_i^{*}\orb_i}{n^{\sigma}} \label{eq:zm:cleverZeroSP} \nonumber\\
 &=& \frac{\partial n^{\sigma}}{\partial \alpha} - \frac{\partial n^{\sigma}}{\partial \alpha} = 0
\end{eqnarray}
This leads directly to:
\begin{eqnarray}
 \sum_i^{N^\sigma} \orb_i\,Q^{i\,*}_x +\orb_i^{*}\,Q^i_x &=& 0 \nonumber\\
 \sum_i^{N^\sigma} \orb_i\,Q^{i\,*}_y +\orb_i^{*}\,Q^i_y &=& 0 \nonumber\\
 \sum_i^{N^\sigma} \orb_i\,Q^{i\,*}_z +\orb_i^{*}\,Q^i_z &=& 0,
\end{eqnarray}
showing that there is no contribution to the potential from the term appearing in the 
functional derivative of an orbital with respect to the density
$\frac{\delta \vec{R}[n(\vec{r})]}{\delta\,n(\vec{r}^\prime)} \cdot 
\sum_{\vec{k}} a^i_{\vec{k}} \, i\,\vec{k}  \,  \phi_{\vec{k}}[n(\vec{r})]$ \\
$= \frac{\delta \vec{R}[n(\vec{r})]}{\delta\,n(\vec{r}^\prime)}\cdot\vec{Q}$.
The remaining four terms used to calculate the potential contribution have the same structure, 
\begin{eqnarray}
&&\frac{1}{2} \sum_{ij} \int  \int {\rm d}\vec{r}_1\, {\rm d}\vec{r}_2 \, U(\vec{r}_1,\vec{r}_2) \nonumber \\
&&\times  \orb_j^{*}(\vec{r}_2)\,\orb_i(\vec{r}_1)\,\orb_j(\vec{r}_2)\,
   \frac{\orb_i^{*}(\vec{r}_1)}{2n^{\sigma}(\vec{r}^{\prime})} 
\,\delta(\vec{r}_1-\vrs) \delta_{\sigma_i,\sigma} \\
&=& \frac{1}{2}\sum_i \delta_{\sigma_i,\sigma} \frac{\orb_i(\vrs) \orb_i^{*}(\vrs)}{2n^{\sigma}(\vrs)} 
    \int \hspace{-.1cm} {\rm d}\vec{r}_2 \, U(\vec{r}_1,\vrs) \sum_j \orb_j(\vec{r}_2) \orb_j^{*}(\vec{r}_2) \nonumber \\
&=&\frac{1}{2} \frac{1}{2} \int {\rm d}\vec{r}_2 \, U(\vrs,\vec{r}_2) \, n(\vec{r}_2) = \frac{1}{4}\, v_{\rm H}(\vrs).
\end{eqnarray}
Due to the product rule during the functional differentiation this term appears four times 
so that the contribution to the potential is exactly the Hartree potential. 
One should note that taking the functional derivative with respect to the 
spin-up or spin-down density leads to the same result.
All the above relationships apply in the same way in the non-spin polarized case.

We have now proven that the proposed procedure to calculate the functional derivative of the Hartree term 
leads exactly to the Hartree potential. %
\section{Experimental Data for the Atom Series}
\label{appendix:exp_data}
In practice, it is not easy and in most cases impossible to measure quantities corresponding to the total energy of calculations. 
In principle, one has to ionize all electrons and measure the energy. The total energy is then the sum of all ionization energies. For the first few atoms of the periodic table this seems to be possible, 
but with larger $Z$ it becomes harder and harder to ionize all of the electrons. A complete set of wavelength corresponding to 
all of the ionization potentials from Hydrogen to Calcium ($Z=20$) has been published~\cite{Moore} quite some time ago. One can now use the 
CODATA~\cite{RevModPhys.80.633,10.1063/1.2844785}, containing the latest values of physical constants and their errors, to convert the measured wave length to energies.
The second column of Table~\ref{table::atomic_energies} labeled EXP$^{(a)}$ contains these data, with the estimated error bars.

Based on the experimental data~\cite{Moore} some effort has been made to try to calculate non-relativistic correlation energies and 
relativistic corrections to the ionization potentials~\cite{PhysRevA.44.7071,PhysRevA.47.3649}.
The data from reference~\cite{PhysRevA.47.3649} seems to be widely used in the literature. But since they were determined by calculations 
rather than measurements, or only partially, we have doubts that these energies can be quoted as experimental data.
Because it is not clear which data set should be used, we decided to quote both in this paper.

\bibliographystyle{unsrt}
\bibliography{pair_density_long_paper}

\end{document}